\documentclass[preprint,12pt,authoryear]{elsarticle}
\usepackage[margin=.7in]{geometry}

\usepackage{amssymb}
\usepackage{amsmath}
\usepackage{natbib}
\usepackage{booktabs}
\usepackage{xcolor}
\usepackage{soul}
\usepackage{float}
\journal{Poetics}
\usepackage[breaklinks=true]{hyperref}
\usepackage{placeins}       
\usepackage{threeparttable}
\usepackage{dcolumn}
\newcolumntype{C}{D{.}{.}{2.5}} 
\newcolumntype{A}{D{.}{.}{2.6}} 

\newlength{\ergmTableWidth}
\begin{document}
\begin{frontmatter}

\title{Follow or Surpass? Collaboration, Sound, and the Recognition of Influence in Lineages of Jazz Trumpet Players}

\author[ceu]{Onur Tuncay Bal}
\author[geneva]{Samuel Coavoux}
\author[ceu]{Nathalia A. Loureiro}
\author[ceu]{Balazs Vedres}
\author[ceu]{Mark Wittek}

\affiliation[ceu]{%
    organization={Central European University, Department of Network and Data Science},
    city={Vienna},
    country={Austria}
}

\affiliation[geneva]{%
    organization={University of Geneva, Geneva School of Social Sciences},
    city={Geneva},
    country={Switzerland}
}

\begin{abstract}
Cultural fields cohere not only horizontally, through collaboration, but also vertically, through lineages of artistic influence. Considerable attention has been devoted to collaboration networks in various fields of creative production; however, we know less about the structure of artistic influence that is nevertheless key to individual artistic development and the evolution of genres. We examine the structure of such an influence network in a genealogy of 52 prominent jazz trumpet players, combining musicians' discographies, demographic characteristics, geographical locations, and audio analysis of more than 16,000 recordings with exponential random graph models. Direct collaboration, overlap in collaborators, and sonic context similarity, that is, the similarity of the recordings on which players appear, are each associated with an ascribed influence tie, net of the others. Not surprisingly, ascribed ties connect musicians whose recordings share a similar sonic context, but musicians also tend to collaborate with their influences and share collaborators with them. However, players whose recordings share a sonic context and who move in the same circle of collaborators can be thought of as peers or rivals, rather than as influences of each other, because lineage implies a succession predicated on difference as much as on continuity. Taken together, our findings suggest that a field's account of its own evolution is neither a record of who played together nor a map of shared sonic contexts, but a selective ordering that requires both. While our data do not permit a causal account, our aim is to provide a rich and varied methodological strategy and descriptive as well as inferential tools to understand the yet understudied vertical integration of cultural fields through influence ties. 
\end{abstract}

% \begin{highlights}
% \item Expert-ascribed jazz lineage tracks both social proximity and sonic context.
% \item Direct collaboration and shared collaborators each explain ascribed lineage.
% \item Where social and sonic context proximity are both high, ascription is less likely.
% \item Pairs sharing a scene and a sonic context are read as peers, not as a lineage.
% \item Recognition of lineage requires sonic context similarity.
% \end{highlights}


\end{frontmatter}

\section{Introduction}
\label{sec1}

\noindent Creativity today is increasingly understood as a collective phenomenon, with collaborative communities emerging in a wide range of fields \citep{becker_art_2008, bourdieu_field_1983, crane_production_1992, crossley_connecting_2019, lutter_creative_2014}. We witness the increasing dominance of teams in science \citep{wuchty_increasing_2007}, applied and fine arts \citep{white_canvases_1993}, music \citep{lena_banding_2012}, and filmmaking \citep{cattani_coreperiphery_2008}. In all of these fields, creators build on each other's ideas as they strive to craft successful products, weave creative careers, and leave a mark. Consequently, a long line of research has examined how collaboration networks affect the creativity and success of individuals or teams \citep{vedres_tension_2026, burt_structural_2004, de_vaan_game_2015, lutter_women_2015}, and how artists influence each other \citep{lena_sincerest_2013, burgdorf_communities_2024, cherifi_social_2017}. Several researchers have studied collaboration in the context of jazz music as well, focusing on network positions that elevate jazz musicians' success \citep{vedres_network_2021, vedres_forbidden_2017}, and how the structure of collaboration relates to genres and styles \citep{kirschbaum_categories_2015, phillips_jazz_2011, phillips_shaping_2013}. 

Cultural fields do not cohere only horizontally through collaboration networks. They also organize their histories vertically through \textit{lineages}: selective accounts of who carried whose work forward.  Practitioners, critics, and educators produce such accounts when they locate artists within a succession and distinguish historically consequential relations from the many sources on which creators draw \citep{bourdieu_field_1983, becker_art_2008, anheier1995forms}. A lineage, in this sense, is a classification \citep{dimaggio_classification_1987, bowker_sorting_2000}: a retrospective and selective representation of descent that a field produces about itself. This article asks what such a classification records. Musicians are connected to one another in many ways: they play together, they draw on the same circles of collaborators, and they record in similar sonic contexts. Which of these proximities does an expert account of descent treat as lineage, and by what logic does it select among them?

To address this question, we take an expert-curated genealogy of jazz trumpet players, with ties we call \textit{expert-ascribed influence relations}, as a starting point for our investigation. We consider two forms of proximity that can each inform ascribed influence: social proximity, captured by direct collaboration and by overlap in collaborators, and proximity in sonic context, captured by audio embeddings of the recordings on which musicians appear. We speak of sonic \textit{context} rather than of a player's sound because the embeddings are computed from complete recordings rather than from isolated trumpet parts: they describe the ensemble, arrangement, and recording within which a player's work was heard, of which the player's own playing is one component. We enrich the genealogy with musicians' discographies, demographic characteristics, geographical locations, and sonic information on over 16,000 recordings, and analyze this combination of data sources using exponential random graph models \citep{lusher_exponential_2013}, which allow us to assess the conditional association of each form of proximity with expert recognition while accounting for the structure of the influence network itself.

Most existing accounts of cultural influence measure it by tracing how artists refer to one another's work --- musicians who sample their peers' records \citep{lena_sincerest_2013}, filmmakers who reference each other's movies \citep{burgdorf_communities_2024}. Such designs infer the outcome from the same class of data as their predictors. Our article instead starts with a judgment by a jazz expert explicitly concerned with influence and takes the relational structure the expert identified as its dependent variable. The genealogy we model was drawn by the jazz trumpeter and educator John McNeil and re-published in Paul F. Berliner's \textit{Thinking in Jazz} \citep{berliner_thinking_2009}, where Berliner records that the figure is ``based on sketch provided by John McNeil, Sept.\ 1993'' (p.~763) and refers readers to McNeil's own discussion of the depicted relationships for their substance \citep[26--27]{mcneil_art_1993}. Our ambition is to explore the correlates of influence ties systematically and measure the extent to which sonic context similarity and collaborative network alignment inform expert-recognized influence.  We do not pretend to causally explain influence, but to shed light on the logic experts use to order a field's tradition into predecessors and successors. 

We combine archival data covering most jazz recordings made in the 20\textsuperscript{th} and 21\textsuperscript{st} centuries \citep{vedres_forbidden_2017, kirschbaum_categories_2015, phillips_shaping_2013} with machine-learning techniques for quantifying the content of recorded music \citep{diwakar2024vggish}. After digitizing and manually enriching McNeil's genealogy with musicians' demographic characteristics, we retrieve their discographies from the Jazz Discography Online \citep{lord_jazz_2020}, yielding more than 5,000 unique recorded sessions and about 18,000 collaborators of the focal musicians. From these sessions, we construct two distinct social measures. The first is \textit{direct collaboration}: whether two genealogy players themselves appeared in a recording session together, a condition of co-presence and direct exposure. The second is \textit{overlap in collaborators}: the share of one player's collaborators that is shared with another, a measure of common position in the wider community rather than of contact. We then obtain audio for more than 16,000 tracks and compute artist-level sonic context profiles from the full recordings using the pre-trained VGGish embedding, widely used in genre classification and recommendation~\citep{hershey2017}, from which we derive pairwise sonic context similarity. Finally, we bring all three measures together in exponential random graph models \citep{lusher_exponential_2013}.

Our results show that direct collaboration, overlap in collaborators, and sonic context similarity are each associated with expert-ascribed influence, and that each holds net of the others: recognized lineage is neither a purely social nor a purely stylistic matter. Almost half of the recorded influence relations join players who actually recorded together, against a far lower base rate among all pairs. However, the association is conditional: sonic context similarity and social proximity act as partial substitutes. Sonic context similarity is most strongly associated with an ascribed influence tie where collaborator overlap is low, and its association weakens as overlap increases, so that pairs close on both dimensions are less likely to be linked than the two associations would jointly imply. Pairs who share a scene and a sonic context are not recorded as a lineage. We read this as a distinction between two relational forms that a canon keeps apart. A cluster of players who collaborate with the same people and record in similar sonic contexts constitutes a scene or a collective pursuing a shared vision; a lineage, by contrast, is a temporal ordering, and to be recognized as one link in it, a musician must be audibly continuous with a predecessor and audibly distinct from them. Whether that distinction is enacted by musicians who differentiate themselves from their sources, or imposed by those whose business it is to classify them, is a question we take up in the Discussion. Taken together, the study contributes to debates at the intersection of the sociology of culture \citep{becker_art_2008, crossley_connecting_2019, dimaggio_classification_1987}, computational social science \citep{salganik_bit_2019, askin_what_2017}, and the study of canon formation in jazz \citep{deveaux_constructing_1991, phillips_shaping_2013, lopes_rise_2002}.

\section{Influence in Creative Fields}

\noindent The idea that cultural production is driven by communities of creators rather than individual geniuses is widely shared among social scientists \citep{bourdieu_field_1983, becker_art_2008, crane_production_1992}. In empirical investigations, the emphasis has shifted from individual creators to collectives \citep{vedres_collaborative_2025, uzzi2005collaboration}, and researchers have extensively studied how collaboration networks contribute to creative success \citep{burt_structural_2004, de_vaan_game_2015, rossman_id_2010, lutter_women_2015}. As a result, a growing stream of studies has investigated the spread of stylistic elements using network analysis in domains such as filmmaking \citep{burgdorf_communities_2024, goldberg_social_2013}, contemporary art \citep{basov2024professional, basov_ambivalence_2020}, and painting \citep{white_canvases_1993}. For music, contemporary studies focus on the interplay between collaboration networks, stylistic orientations, and success in punk \citep{crossley_man_2009}, classical music \citep{mcandrew_music_2015}, rap \citep{lena_sincerest_2013, halgin2020prismatic}, dance music \citep{wilderom2019intersecting, wilderom2025dance}, and jazz \citep{vedres_forbidden_2017, kirschbaum_categories_2015}. A shared emphasis of these accounts is collaboration as a horizontal, symmetrical relation, while influence ties are largely sidelined.

% Paragraph 2. Research gap  [sharpened]
Even studies that do engage with influence rarely measure it directly and tend to neglect recognized influence. Instead, influence is inferred from the transmission of behaviors or stylistic orientations through networks of collaboration or affiliation, overlooking intangible aspects of the relation and registering only the manifest dissemination of tangible symbolic elements. While tracing how artists copy stylistic elements yields important insights, the question of how cultural fields organize into lineages of influential players deserves attention in its own right. Experts and field participants construct lineages to make sense of a field's history, communicate its achievements to a public audience, and to guide the education of new entrants. Scholars in adjacent fields use ethnographic data and interviews to study the full bandwidth of influence among scientific \citep{latour_laboratory_1986} and cultural producers \citep{faulkner__2009, berliner_thinking_2009}, but such qualitative approaches rarely apply network analysis to the organizing principles of influence relations. Meanwhile, a growing stream of research quantifies the content of cultural fields, for instance by applying audio analysis to the success of popular music \citep{askin_what_2017, kim2024feature} or by analyzing the color-coding of album covers across genres \citep{sgourev2023relations}. A network-analytic perspective is usually absent from this work. Here, we build on insights from prior network research on cultural influence \citep{cattani_coreperiphery_2008, burgdorf_communities_2024, crossley_man_2009, mcandrew_music_2015}, qualitative accounts of influence among artists \citep{becker_art_2008, baumann_hollywood_2007, berliner_thinking_2009}, and work that quantifies the content of cultural fields \citep{kim2024feature, askin_what_2017, mohr_measuring_2020}.

\subsection{Tangible and intangible influence}
We see two crucial differences between recognized influence and collaboration. First, in contrast to collaboration, recognized influence relations are predominantly asymmetrical, and the asymmetry is often paired with a distance in space and time, sometimes to the extent that there is little overlap in the natural lives of the artist and their influence. We do not claim that mutual influence is absent from creative fields; peer and competitive influence are well documented, and our own data contain reciprocated pairs. Our focus is on lineage-type asymmetric relations, one well-attested form among others. Creators often intentionally draw on prior eras for generative influences \citep{uzzi_atypical_2013} and publicly acknowledge them, featuring stylistic elements that signal their belonging to part of the jazz tradition \citep{murphy1990jazz}. Second, influence includes the transmission of both tangible and intangible knowledge, and this is what leads us to expect that sonic context similarity and social proximity carry different information about ascription.

Influence in the narrow sense is a channel of transmission that includes the circulation of explicit techniques, themes, and approaches closely relevant to the cultural product. In music, the most direct example is when an artist copies a section of audio from another's recording and inserts it into their own, a practice prevalent in hip hop and rap \citep{lena_sincerest_2013, demers_sampling_2003}. In jazz, musicians learn a canon of standards that enables them to play together in jam sessions \citep{faulkner__2009}, and it is well documented that they explicitly borrow particular motifs, melodies, and ``licks'' from their influences \citep{faulkner__2009, kirschbaum_categories_2015}. Musicians emphasize engaging with the solos of favored players to develop their own improvisational style \citep{niles2009pat, priestley2007chasin, davis1990miles, porter1999john}. \citet{berliner_thinking_2009} records the transmission of a solo repertory as an ordinary social transaction among trumpet players, and one that participants themselves understood as lineage: Charlie Shavers, before teaching Dizzy Gillespie the solos he had memorized from Roy Eldridge recordings, complained, ``Why don't you go steal your own shit, instead of getting Roy secondhand?'' (p.~96). Sonic resemblance is thus the most immediately available signal from which influence can be ascribed, since it requires no knowledge of who was in the studio; one need only listen. It is also the signal that musicological analysis tends to focus on, as Berliner's own attention to air in the tone, tonguing, and articulation illustrates, and the signal on which the genealogy we model is explicitly built, organized as it is around ``the two main schools of post-war jazz trumpet sound production'' \citep[25]{mcneil_art_1999}. Sonic resemblance is not, however, a sufficient index of influence even on the source's own account: McNeil describes Thad Jones and Clark Terry as players ``whose personalities are strong enough to disguise their influences,'' so that influence may be present in a career and not audible in it --- a possibility we return to in the Limitations. A further qualification concerns what can be measured. On record, a trumpeter's sound is embedded in an ensemble, an arrangement, and a recording; the computational measure we use describes this recorded \textit{sonic context} as a whole rather than the trumpet part in isolation (Section~\ref{sec:soniccontext}).

Social proximity carries different information, and in two distinct forms. \textit{Direct collaboration} is a condition of co-presence. For trumpet players it typically means a big band or a larger ensemble with several horns, and such ensembles have long functioned as informal educational institutions in which younger musicians developed their technique and learned directly from older members of the community \citep{berliner_thinking_2009, faulkner__2009}. Co-performance provides the setting in which tacit elements are transmitted that are not audible on recordings: aesthetic judgments, bandstand conventions, and ways of relating to other players. Artists borrow from their influences a manner of being in the field, not only a manner of playing \citep{bourdieu_field_1983}; jazz musicians learn with whom to play and how to respond to what others play by observing those they admire \citep{berliner_thinking_2009}. Different kinds of collaborative ties may carry different implications here, since roles within ensembles are differentiated and formal and network leadership is distinct \citep{van2023network}. 

\textit{Overlap in collaborators}, by contrast, is not a condition of contact but of common position in the community of musicians. Two players who draw on the same wider circle of musicians belong to the same scene: they are pursuing a related vision of how the music should sound, frequently in the same city and often within the same generation, and frequently at the center of the community, where a core develops that determines where the music goes at a given time \citep{phillips_shaping_2013, crossley_man_2009, lena_banding_2012}. Scripts for how music should sound---which chords, which rhythms, which instrumentation ---travel through these shared personnel. Both conditions should therefore be legible to an expert, yet in different ways: co-presence as evidence of contact, and shared circles as evidence of common location in the field. 

These considerations lead to our first research question (\textbf{RQ1}): Do sonic context similarity, direct collaboration, and overlap in collaborators each contribute independently to the probability that an influence relation is recorded between two players, or is one largely a proxy for another? With this research question, we first consider the more obvious sonic dimension that any jazz listener would think of at first as a factor in influence, which we measure as similarity in sonic context. When listening to a trumpet player, one is likely to think of key influences with similar sound and phrasing. However, we also take into account that musical influence would leave its traces in the network structures among musicians, which listeners cannot hear and that are more hidden but nevertheless important. 

\subsection{Following and surpassing idols}
Somewhat obviously, if a musician is influenced by another, they will sound similar, but how similar? At the extreme, a musician could be a clone of another. In other words, how far can we go along the dimensions of sonic context and social proximity in expecting an increase in the likelihood of an influence tie? If influence is about being artistically faithful, at the extreme, influence would entail complete assimilation; it would stifle innovation, and fields would exhibit stasis. Critics and scholars of cultural fields readily acknowledge that many artists are imitators. The trumpet player Art Farmer put it bluntly:
\begin{quote}``Basically, ninety-nine and nine-tenths percent of everybody out there is just copying somebody else. Here in New York, I remember every piano player was trying to play like Horace Silver at one time, and then later on, everybody was trying to play like Bill Evans. Some of the guys who were playing like Horace a couple of years later were trying to play like Bill. And then everybody was trying to play like McCoy Tyner.'' \citep{berliner_thinking_2009}\end{quote}

Yet cultural fields do change, and individual artists strive to make contributions that surpass their influences \citep{white_canvases_1993, bourdieu_field_1983, becker_art_2008}. Influence in this sense entails a productive tension: it is simultaneously about learning existing vocabularies \textit{and} achieving differentiation from them. Musicians describe the tension in exactly these terms. Lonnie Hillyer, reflecting on his development, told \citet{berliner_thinking_2009}: ``When I was younger it was like I was trying to get what Dizzy felt through the horn, but, eventually, it's got to come to be about you'' (p.~261). Charlie Parker's injunction to ``learn the changes, then forget them'' \citep{pugatch2007acting} makes the same point, as does John Coltrane's conviction that only immersion in the work of past masters permits advance to novel expression \citep{porter1999john}.

This tension resonates with non-monotonic patterns observed in adjacent domains: songs that are too typical or too atypical underperform on the charts \citep{askin_what_2017}; genre movements that fail to differentiate from their predecessors are reabsorbed, while those that differentiate too sharply lose recognizability \citep{lena_classification_2008}; visual artists are expected to produce work that is both true to their past selves and sufficiently novel \citep{wohl2019_CreativeVisionsPresenting}; and in poetry, influence has been read as both borrowing and a necessary misreading that establishes originality \citep{bloom_anxiety_1997}. We stress, however, that these accounts motivate the shape of a relationship and not the location of any particular boundary. Neither they nor our argument implies a determinate threshold, so we treat the following question as exploratory rather than confirmatory: we estimate the top-decile indicator alongside the linear term, so that it captures a departure from linearity at the upper end of the distribution, and report a sensitivity analysis across cutoffs.

Similarly to extreme sonic context similarity, one can think about the extreme case in social proximity when a musician would borrow or inherit the exact same set of musicians that his or her influence played with. Too much proximity, however, signals that two players are colleagues or competitors occupying a similar place in the jazz community, which entails playing in the same clubs and recording with the same personnel. These considerations lead to our second research question (RQ2): is the association between sonic context similarity as well as social similarity and influence non-monotonic --- specifically, does it weaken at the upper end of the observed similarity distribution?

\subsection{Scenes and lineages}
If sonic context and collaboration proximity each carry information about the presence of a lineage influence tie, what happens when they coincide? One possibility is complementarity: if collaboration and sonic context capture distinct facets of a transmission relation, each should contribute additively, and their joint presence should reinforce the probability of an ascribed relation. Field theory holds that influence is a relation between \textit{distinct} positions: it presupposes that the master occupies a place the influenced artist is moving toward or away from \citep{bourdieu_field_1993, white_canvases_1993}. Two musicians who share a collaboration neighborhood and a sonic context occupy the same position in the field. They collaborate with the same alters, which is to say they are located in the same region of the community, and their recordings sound alike, which is to say they are to some extent exchangeable. Structurally equivalent actors of this kind are peers, competitors, or co-members of a project, rather than being influence ties. An influence tie encodes progression or succession. For an influence relation to be legible, a musician must go a step beyond their predecessor, establishing a recognizable voice and network position of their own while building on elements taken from those who came before.

The distinction, in short, is between two relational forms that structure cultural fields. A cluster of players who are close both in sonic context and socially can be defined as a \textit{scene}--- a collective pursuing a shared artistic vision at a given moment. A \textit{lineage} is a different relation: diachronic rather than synchronic, ordered rather than clustered, and defined by difference as much as by continuity. We argue that a canon of lineage records the second and not the first, and we test that argument by asking whether the joint presence of social proximity and sonic context similarity is associated with a \textit{lower} probability of an ascribed relation rather than a higher one. The reasoning is straightforward once the two dimensions are held apart. To be close on both is to occupy the same position in the field. To be close on one and not the other is to stand in a relation that can be read as descent: a player who takes up a predecessor's sound while collaborating with a different circle of musicians, or who inherits the same collaborators while pursuing a sound of their own. 

Two readings of such a pattern are available to us, and we flag now that our design cannot adjudicate between them. The first is substantive: musicians who are socially close differentiate themselves aesthetically, because a field that prizes novelty rewards those who find their own sound, so that pairs who remain indistinguishable are pairs in which the succession did not occur. The second is classificatory: experts define lineage so as to exclude close resemblance, reserving the label for musicians perceived as carrying a tradition forward rather than reproducing it, so that the pattern describes the concept of influence in use rather than the conduct of musicians. The second reading has some direct support in our source, whose chart is organized around a theory of sound production and whose author was therefore demonstrably reasoning in aesthetic terms when deciding who belongs where. It also has a counter-case there: McNeil does not treat close resemblance as disqualifying, describing Jon Faddis as ``a clone of Dizzy Gillespie stylistically'' and placing him as close to Dizzy as the diagram allows rather than omitting him \citep[25]{mcneil_art_1999} --- although Faddis does not appear in the version of the genealogy we model. Both readings are consistent with the same coefficient. We return to them in the Discussion rather than building either into the question.

It is relevant here that the two dimensions are not equally malleable. Musical sound, and the sonic context in which it is recorded, can be altered; collaborative networks are stickier, built through years or decades of trusted contacts, and artists do not readily change their embeddedness in them. Musicians sometimes recruit a new player to change the sound of an established band, as Miles Davis did in bringing the pianist Wynton Kelly into his group; asked what Kelly contributed, Davis answered that ``Wynton's the light for a cigarette. He lights the fire, and he keeps it going. Without him there's no smoking.''\footnote{Miles Davis, quoted by Nat Hentoff in the liner notes to Wynton Kelly, \textit{Kelly at Midnite} (Vee-Jay VJLP-3011, 1960). We cite the passage at second hand, from \textit{JazzProfiles}, \url{https://jazzprofiles.blogspot.com/2019/10/wynton-kelly-1931-1971-pure-spirit.html}.} More often, however, musicians remain in the same collaborative context while striving for a new sound. Davis was a notably candid bandleader about conjuring a new sound from the same set of collaborators by challenging their routines.
\begin{quote}``See, if you put a musician in a place where he has to do something different from what he does all the time, [...] he'll be freer, will expect things differently, will anticipate and know something different is coming down. I've always told the musicians in my band to play what they know and then play above that. Because then anything can happen, and that's where great art and music happens.'' \citep[p.~220]{davis1990miles}\end{quote}
Differentiation, in other words, is pursued along the sonic dimension while social position is held relatively constant --- which is what makes the conjunction of the two a meaningful configuration rather than a redundancy.

Taken together, these arguments lead us to our third research question (RQ3): is the joint occurrence of high social proximity and high sonic context similarity associated with a lower probability of an ascribed influence relation than either condition alone would imply?

\section{Data}

This section describes the data on which our analysis rests and the procedures by which we assembled them. The design combines a small, expert-curated dependent network with large-scale archival and audio data on each of its members. The dependent network is a genealogy of 52 jazz trumpet players connected by 74 ascribed influence ties. To characterize the relations between these players, we draw on three further types of data: recording sessions, which document who played with whom, when, and where; audio recordings of those sessions, from which we derive sonic context profiles; and biographical and release information, which situates each player in time and captures their success. Although the genealogy itself is small, each of its members is represented by an extensive record of recording activity, so that the measures we construct rest on a substantial empirical base.

Because the genealogy records McNeil's judgment as of September 1993, recordings and collaborations that took place after that year could not have informed it. We therefore restrict all session-based data to sessions recorded up to and including 1993. The restriction applies to the collaboration networks, to the recording locations and years from which spatial and temporal overlap are computed, to the audio from which sonic context profiles are derived, and to the release counts used to measure success; birth years are the only covariates it leaves unaffected. Sessions are dated by recording date rather than release date, so that later reissues of earlier sessions are retained and recent compilations do not enter the data. As a consequence, each player is represented by their career as it stood when the genealogy was drawn, and the youngest players in the genealogy are represented by comparatively short recorded careers.

\subsection{Jazz as a case}
Jazz is a suitable domain for understanding how lineage is recognized in a creative field, first because of its culture of promiscuous relationality: jazz musicians tend to play with a wide variety of peers throughout their careers \citep{berliner_thinking_2009}. Compared with neighboring fields such as rock or classical music, jazz shows a much lower level of fixity to collectives, bands, or orchestras \citep{lena_banding_2012}, and is characterized by a project-based, fluid social structure with an emphasis on musicians' individual artistic vision, where the fixed entity is not so much the band as the band leader \citep{phillips_jazz_2011, kirschbaum_categories_2015}. Collaboration networks are accordingly an essential conduit for the circulation of ideas and musical material, and recording sessions allow us to infer the community’s social structure. Second, jazz combines a strong tradition of shared conventions with an equally strong ethos of innovation, so that the question of who stands in whose line is one the field itself continually poses. Third, jazz is a field in which lineage matters to audiences and participants: discussing who influenced whom and which strands emerge in the jazz tradition interests critics, aspiring musicians, jazz historians, musicologists, and listeners alike \citep{deveaux_constructing_1991, lopes_rise_2002}. These properties make it a strategic research site for studying the ascription of influence, the formation of tradition, and the emergence of innovation.

\subsection{Lineage as expert recognition}

Here, we focus on \textit{ascribed influence}: a directed tie recorded in an expert's judgment of who influenced whom. Consequently, we study the structure of an expert's representation of influence ties rather than self-reported influence or influence inferred from behavioral traces. The chart externalizes a classificatory judgment by selecting certain artists and relations as historically meaningful. Mental models like this guide jazz education and help jazz critics and journalists legitimize jazz as an art form with its own history by describing musicians as exponents of critical musical developments \citep{deveaux_constructing_1991, phillips_shaping_2013, lena_banding_2012}. Constructing a lineage thereby supports a music's collective identity by giving practitioners a shared set of reference points. Crucially, it supports the training of new musicians, which is precisely the setting in which the genealogy we analyze was produced: McNeil originally drew it for a book that teaches the art of jazz trumpet \citep{mcneil_art_1993}. 

It's worth discussing that expert ascription and self-ascription are distinct forms of cultural judgment. When musicians name their influences, they present a biographical account of sources they regard as important to their personal creative development. Such accounts may also position the musician within a tradition and can be selective or strategic. An expert genealogy follows from different aims: it compares artists across careers and selects relations that fit an account of how a style or practice developed. When an expert ascribes influence, they do so at greater distance and in hindsight, assessing a musician's contribution in terms of how it shaped the field; the account is retrospective, comparative, and selective, and it registers developments over a longer horizon. The two forms of ascription may overlap, but they need not select the same predecessors or assign them the same historical significance. For example, Tommy Turrentine, interviewed by \citet{berliner_thinking_2009}, names seven players he learned from---``Fats Navarro, I loved him; Art Farmer, I listened to him. Miles, Dizzy, Woody [Shaw], Donald Byrd, Blue Mitchell'' (pp.~138--139)---six of whom appear in the genealogy, yet McNeil decided to see Turrentine as being influenced by Clifford Brown, a judgment consistent with perceived similarities in phrasing and articulation. Taken together, expert-ascribed influence is more selective than self-reported influence, as it distills a musician’s position in the lineage through a relational judgment informed by a close musicological reading of the field.

The genealogy we analyze illustrates each of these properties and the compromises they entail. McNeil drew it for a book teaching the art of jazz trumpet \citep{mcneil_art_1993, mcneil_art_1999}, and he is explicit that its purpose is to order the field by picking key exponents and connecting them in a lineage rather than archiving all influences a player might have had. The rule is stated outright in McNeil's book: ``One early representative from each school is enough'' \citep[25]{mcneil_art_1999}. Inclusion is thus governed partly by redundancy: a player whose contribution is already represented by someone else is not included, however real their influence. For instance, Freddie Webster is left out although ``many trumpet players claim to have been influenced by him,'' because without him ``the two-school theory of jazz trumpet sound production would have been unaffected, and that, more or less, is the whole point of the chart'' \citep[25]{mcneil_art_1999}. Other choices were ``based mostly on the need to compress the chart into a reasonable space and avoid crowding the chart with too many influence ties''. Consequently, McNeil aimed to offer a distilled picture of the tradition of jazz trumpet playing rather than faithfully collecting musicians' self-reported influences. The decision of who is included in the lineage is a part of the experts' mental model and shaped by a field's opportunity structure as much as its aesthetics; we return to this point in the Limitations.

Independent of the considerations behind its construction, the genealogy can be described as a directed network whose structural properties we summarize below. It comprises 52 players and 74 ties and is sparse (density $= 0.028$; mean in- and out-degree $= 1.42$). Ties are predominantly asymmetric: the dyad census  records 2 mutual, 70 asymmetric, and 1{,}254 null dyads.\footnote{Note that 18 players have two or more predecessors. Hence, the genealogy is not a tree, although visual inspection suggests a tree-like structure. Additionally, since two dyads are mutual, it is not a directed acyclic graph either.} The network is also largely ordered in time, with most ties running from an older to a younger player. Most players are assigned a single source of influence: 31 have exactly one predecessor. The out-degree distribution, by contrast, is highly unequal. The majority of players have no outgoing ties, while outgoing ties concentrate in a few figures, such as Freddie Hubbard, Clifford Brown, Miles Davis, and Dizzy Gillespie. This concentration may reflect these players' historical prominence, as well as canon formation and McNeil's selection of eminent figures. Closure is rare, and most ties have no edgewise shared partner. Full descriptives are reported in Table~\ref{tab:netdescriptives} and~\ref{app:structure}.

In summary, constructing a genealogy involves more than identifying influence relations. It requires selecting exemplary figures, identifying key influence relations among a large set of possible influences to arrive at a legible representation, and arranging artists according to a theory of stylistic development. Therefore, we see the resulting network not as an incomplete measurement of an underlying influence network, but as a deliberately selective representation that expresses an expert account of what counts as consequential development in jazz history.

\subsection{Collaboration Data}
We retrieve recording sessions from \textit{The Jazz Discography}. The unit of observation for this dataset is a session. Each session documents an instance in which musicians performed together to record a set of tracks. These sessions allow us to identify the collaboration networks of genealogy players because each session entry contains detailed information about who participated and in which role. This lets us identify the musicians involved in a session, the location and date, and the recording label codes (catalog numbers) for releases. The dataset is manually curated and continuously updated through careful verification. It contains about $1{,}5$ million artists and more than quarter of a million sessions, and is regarded as the most comprehensive jazz database for research, with documented use across several disciplines \citep{vedres_forbidden_2017, proutskova2022jazz, hannibal2015network, phillips2009pseudonyms}. It is important to note that while the sample size of the genealogy is limited ($N = 52$), the information we used to augment the data has a substantial scale. In particular, we used the digital archive to retrieve all sessions of genealogy players, yielding more than 5,000 sessions and 62,000 individual tracks. Subsequently, we obtained all co-musicians of the genealogy players, yielding a network of over $18{,}000$ musicians. 

\subsection{Audio Data and Audio Embeddings}
Although retrieving session metadata is relatively straightforward, obtaining the \textit{audio recordings} for those sessions is considerably more difficult for several reasons. First, a single session is often distributed across multiple releases, frequently in fragmented or inconsistent ways. For example, the track \textit{Pluckin' the Bass} by Chu Berry with Cab Calloway---featuring Dizzy Gillespie on trumpet---appears on a collaborative LP titled \textit{Dentistry in Rhythm -- Volume 3}. However, the same title was also used by another company for a different compilation containing a different selection of tracks. The fact that some recordings are released multiple times creates a disambiguation issue: it's non-trivial to map recording sessions to releases, since multiple releases may or may not include the required recording.

Secondly, searching for track names directly via a streaming platform’s API rarely returns the specific session version we are looking for. Jazz standards are performed repeatedly across many sessions, which introduces substantial ambiguity. For instance, Chet Baker recorded \textit{I Fall in Love Too Easily} multiple times in different sessions. Owing to the improvisational nature of jazz, these recordings can differ significantly in their sonic characteristics.

To address these challenges, we employ a highly selective processing pipeline. First, we retrieve the session information and identify the albums in which the corresponding tracks were released by using their label codes. Because there is no unified database with API access, we rely primarily on Discogs, where we search for each album’s label code. If a session has been released across multiple formats, we query all associated label codes. Second, after retrieving the album data, we match tracks across releases using a fuzzy string-matching algorithm with \textit{RapidFuzz} \citep{max_bachmann_2025_15133267}. Once a track-album match is confirmed based on the fuzzy-matching score ($.98$), we extract the International Standard Recording Code (ISRC) for that track. The ISRC is a unique identifier assigned to individual audio recordings, enabling us to distinguish between different versions of the same piece of music. This procedure leads us to find the exact recording of that track in that exact session. 

Once the ISRC codes are obtained, we query the Deezer API using these unique identifiers. Because ISRCs correspond to specific recordings rather than song titles, this approach allows us to reliably retrieve the correct 30-second audio preview for each track. These previews are not fixed to a particular segment of the song; instead, they provide a continuous, uncut 30-second excerpt starting at a random timestamp in the song. As a result, these excerpts offer an unselected snapshot of the recorded performance as a whole, including the ensemble, the arrangement, and the recording, within which the focal musician's playing is one component. They capture sufficient musical information for our downstream analytical tasks, consistent with prior applications \citep{ahmed2024musical, tzanetakis2002musical, naidu2025classification}. The final dataset comprises approximately 16,000 CD-quality WAV previews, recorded at a 44.1 kHz sampling rate and 16-bit resolution. Figure~\ref{app:coverage} shows the temporal coverage of embedded sessions by artist.

\subsection{Player Demographics and Session Locations}
In addition to the audio samples, we also extract detailed information about the sessions' locations. We first apply a \href{https://huggingface.co/Jean-Baptiste/roberta-large-ner-english}{named entity recognition (NER) model} to identify location entities mentioned in the session metadata. These extracted location strings are then sent to the Google Places API to obtain their corresponding latitude and longitude coordinates. Because automated geocoding can occasionally misinterpret musically specific or historically unusual location names, we manually verify each retrieved coordinate to ensure accuracy. For example, for some recording sessions, the Jazz Discography lists the recording location as ``SS Norway,'' referring to a cruise ship active in the Caribbean; however, the Google Places API incorrectly resolves this as the country ``Norway.'' Our manual checking of the automatically retrieved locations ensures that such cases are corrected.

Finally, we manually collect demographic information for each trumpet player, along with data on their recording output by querying TomLord and Discogs. The demographic attributes include each artist's birth year and, when applicable, death year. Using information from the Jazz Discography, we constructed a variable capturing each artist's success by calculating the total number of albums issued. We exclude posthumous compilations or reissues released long after the artist’s career had ended. This restriction allows us to more accurately approximate the musician’s active creative period. Together, these demographic and success measures provide essential contextual features for linking musical style to biographical background

\section{Measures}

\subsection{Overlap in Collaboration}
We use \textit{The Jazz Discography} to construct the collaboration network among all musicians who played with the genealogy players. To this end, we create a square matrix in which rows and columns index musicians, and a non-zero entry indicates that the corresponding musicians performed together in at least one recording session. We consider all sessions featuring the trumpet players listed in our expert-curated genealogy, along with those involving any of their collaborators. This yields an initial set of $165{,}439$ unique artists. We then reduce the matrix's size by removing all artists who never appeared in a session with any of the genealogy players. The resulting sparse binary collaboration matrix has dimensions $18{,}312 \times 18{,}312$, containing $6{,}682{,}500$ non-zero entries (representing pairwise collaborations).

Subsequently, we derive the \textbf{Shared Collaborators Matrix}. In this matrix, a non-zero entry between two trumpet players indicates that they share at least one common collaborator. To capture how strongly the sets of collaborators overlap between two genealogy players, we calculate the proportion of the target player's collaborators that are shared with the source player. For instance, if Player A played with Ron Carter, Chick Corea, and Herbie Hancock, and Player B played with Ron Carter and Bill Frisell, the proportion of shared collaborators would be 1/3 for Player A and 1/2 for Player B. Hence, we normalize for the volume of collaborators musicians accumulated during their careers. 

Formally, let $\mathbf{S^\text{collab}} \in [0,1]^{N \times N}$ denote the Shared Collaborators Matrix, where the indices $1 \dots N$ correspond to the set of trumpet players. For any two artists $i,j$, let $\Gamma_i$ denote the set of collaborators (non-zero indices in the $i$-th row of the full collaboration matrix). The entry $S_{ij}$ is defined as the directed overlap:
\begin{align*}
    S^{\text{collab}}_{ij} = \frac{|\Gamma_i \cap \Gamma_j|}{|\Gamma_j|}
\end{align*}

This formulation yields an asymmetric matrix in which $S^{\text{collab}}_{ij}$ denotes the extent to which artist $j$'s circle of collaborators is contained within artist $i$'s circle. 

\subsection{Direct Collaboration}

We further quantify the \textbf{direct collaboration} between the genealogy players, that is, whether two trumpet players appeared together in at least one recording session. Using the same session data from \textit{The Jazz Discography}, we restrict the full collaboration matrix to the set of genealogy players, so that both rows and columns index the trumpet players in our expert-curated genealogy.

Formally, let $\mathbf{S^{\text{collab-direct}}} \in \{0,1\}^{N \times N}$ denote the Direct Collaboration Matrix, where the indices $1 \dots N$ correspond to the trumpet players. For any two artists $i,j$, let $\Gamma_i$ again denote the set of collaborators of artist $i$ (the non-zero indices in the $i$-th row of the full collaboration matrix). The entry $S^{\text{collab-direct}}_{ij}$ is defined as
\begin{align*}
    S^{\text{collab-direct}}_{ij} =
    \begin{cases}
        1, & \text{if } j \in \Gamma_i \\
        0, & \text{otherwise}
    \end{cases}
\end{align*}

Since co-presence in a session is a mutual relation, this formulation yields a symmetric matrix, $S^{\text{collab-direct}}_{ij} = S^{\text{collab-direct}}_{ji}$, encoding an undirected network.

\subsection{Audio Embeddings and Sonic Context Similarity}
\label{sec:soniccontext}

Starting from ca.\ 16{,}000 previews in WAV format, we extract audio
representations using \href{https://github.com/tensorflow/models/tree/master/research/audioset/vggish}{\textit{VGGish}}
\citep{hershey2017cnn}, a VGG-like convolutional architecture that maps
$0.96$\,s log mel-spectrogram patches to $128$-dimensional embeddings. The
released checkpoint we employ is the model used to produce the $128$-dimensional
embeddings distributed with the initial AudioSet release
\citep{gemmeke2017audio}, trained on a large YouTube corpus that later became
YouTube-8M. Hershey et al.\ report that embeddings from this class of model
substantially outperform raw spectral features on the AudioSet acoustic event
detection task \citep{hershey2017cnn}.

We adopt VGGish for three reasons. First, our corpus consists of short,
heterogeneous, and unlabeled previews spanning several decades of recording
practice; transfer from a large pre-trained model avoids the annotated training
data that a domain-specific model would require. Second, although VGGish is
pre-trained on general audio events rather than on jazz, its embeddings transfer
to music-specific tasks: they exceed hand-crafted MFCC baselines in music
emotion recognition despite their low dimensionality \citep{koh2021comparison},
serve as a standard benchmark representation in musical instrument
classification \citep{ding2023audio}, and match hand-crafted MIR feature sets in
genre-oriented classification of film soundtracks \citep{ma2021computational}.
Third, within a single corpus VGGish performs comparably to alternative
general-purpose embeddings such as OpenL3 and YAMNet
\citep{wang2023transfer}; since all our comparisons are internal to one
discographic corpus, the cross-domain generalization gap that distinguishes
these models is not the binding constraint here. We note that more recent
self-supervised representations outperform VGGish on genre benchmarks, and we do not claim optimality, only adequacy for the relative comparisons we draw.

These embeddings require a precise interpretation. Each preview is an excerpt of the complete, mixed recording as released, and we apply no source separation to isolate the trumpet from the other instruments. An embedding therefore summarizes the sound of the recording as a whole: the size and instrumentation of the ensemble, the arrangement and repertoire, tempo and rhythmic texture, the acoustic and production characteristics of the recording, and, as one component among these, the focal player's own performance. We refer to this composite as the \textit{sonic context} of a player's work. Aggregated across the recordings on which a player appears, it describes the recorded sound within which that player's work was heard. \textit{Sonic context similarity} between two players accordingly measures how alike the recordings are on which each of them appears; it does not measure how alike their trumpet playing is, and we do not interpret it as such. The descriptive projection in Section~\ref{sec:descriptive} is consistent with this reading: its leading components appear to separate vocal-led tracks from small-combo bebop, early from later recording practice, and big-band from smaller-ensemble recordings.

We nevertheless regard sonic context as a substantively meaningful dimension for the question we pose, for three reasons. First, the sonic context of a recording is in part a product of musicians' own choices. In a field in which the fixed entity is the bandleader rather than the band \citep{phillips_jazz_2011, kirschbaum_categories_2015}, ensemble format, personnel, repertoire, and arrangement are to a considerable extent chosen by the musicians involved, as Miles Davis's recruitment of Wynton Kelly to change the sound of his group illustrates; the context in which a player records is thus in part an expression of their artistic orientation rather than wholly external to it. Second, the transmission discussed above is not confined to individual tone: scripts for which chords, rhythms, and instrumentation a music should have travel through shared personnel, and these are properties of a recording rather than of a single instrumental part. Third, sonic context is the form in which a player's sound circulates on record. Musicians learn from many of their influences through recordings, as Berliner's account of players memorizing recorded solos documents \citep{berliner_thinking_2009}, and on a recording a trumpeter is heard within an ensemble and a production. The focal player's contribution is thus part of the measured mixture rather than excluded from it, although its weight varies with ensemble size and with whether the player is featured in the sampled excerpt.

The same construction sets the limits of the measure. It cannot separate a player's tone, articulation, or phrasing from those of the musicians around them, and two players may have similar sonic contexts because they recorded in the same era, with similar recording technology, or from the same repertoire, rather than because they share a musical orientation. We address recording era with the birth-year and temporal-overlap terms and with a specification that omits them (~\ref{app:no_time}), and we return to the remaining alternatives in the Limitations.

For each artist, we compute the mean of all associated $128$-dimensional embedding vectors, yielding a single representative embedding per artist, which we refer to as the artist's sonic context profile. A cross-sectional dyadic model requires one value per node, so this aggregation is a property of the design rather than a choice among alternatives: within-career stylistic development can only be represented in a specification that models the influence network over time, which the size of the genealogy does not permit. Because each preview samples a recording at a random point, an individual
excerpt need not capture the focal trumpeter. We therefore assess whether the
artist-level mean is stable across random subsets of the available audio; the
procedure and results are reported in~\ref{app:splithalf}.

We define the \textbf{Sonic Context Similarity Matrix}, denoted as
$\mathbf{S}^{\text{sonic}} \in [0,1]^{N \times N}$. The entry
$S_{ij}^{\text{sonic}}$ captures the similarity between the sonic context profiles of artist
$i$ and artist $j$, calculated as the rescaled cosine similarity between their
respective mean embedding vectors:
\begin{equation*}
    S_{ij}^{\text{sonic}} = \frac{1}{2}\left(1 + \frac{\mathbf{e}_i \cdot \mathbf{e}_j}{\lVert \mathbf{e}_i \rVert \, \lVert \mathbf{e}_j \rVert}\right)
\end{equation*}

No session for which audio was retrieved includes two genealogy players, so no preview enters the profile of more than one artist. Across the 2,652 ordered dyads, sonic context similarity and collaborator overlap are moderately correlated ($r = 0.41$). The association is inflated by dyads involving players whose recordings date from the early decades of jazz, whose sonic contexts are distant from those of the other players and who share few or no collaborators with them.

\subsection{Success}

Being successful among audiences and critics is crucial for artists' standing in their cultural field and likely increases their influence on other artists \citep{bourdieu_field_1993}. In our case, famous trumpet players also display more recognized influence ties. For example, Miles Davis, Freddie Hubbard, Dizzy Gillespie, and Clifford Brown are the most influential players according to the genealogy and were highly successful among musicians, critics, and the broader public. We use the number of releases per genealogy player to capture success and control for it in our inferential statistical models. Note that the number of releases is different from the number of sessions. While a recording session indicates that musicians came together in a studio or during a live performance to record a number of tracks, those tracks can be released multiple times and in different constellations. For example, if a musician becomes highly popular, labels tend to release compilations that recombine successful tracks from multiple albums (``best ofs''). Hence, releases capture the behavior of record companies and their responses to audience demand for particular tracks, which in turn reflects the success of musicians. 

\subsection{Demographic Information, Location and Activity Overlap}
 
To account for additional factors structuring the anatomy of influence ties, we included control variables for trumpet players' generational status and the locations where they were active. By considering how close players are in time and space, we account for the fact that influence is more likely when players belong to the same generation and community of musicians within a given region. Including this information is also important to dissect whether collaboration and sonic context similarity still show a relationship with ascribed influence ties once we account for the part of social and sonic context similarity that coincides with being active in similar cities and during a similar time.
 
To analyze the geographical overlap between artists, we utilized the $k$-medoids clustering algorithm with $k=8$ clusters. This method was chosen for its robustness to outliers compared to $k$-means, a property particularly beneficial for geographical data \citep{sadeghi2025clustering, arora2016analysis}.
 
We set $k=8$ on the basis of the resulting partition's substantive coherence and its stability across alternative values. The eight clusters separate the northeastern corridor (New York, Philadelphia, Boston) from the midwestern and southern United States (Chicago, Detroit, Kansas City, St.\ Louis, New Orleans) and from the West Coast (Los Angeles, San Francisco), and further distinguish Western Europe, Japan, South America, Africa, and Australia. The separation of the northeastern from the midwestern cluster matters for our purposes, since these correspond to historically distinct recording scenes. The partition nonetheless remains coarser than the city level (Chicago and New Orleans, for instance, fall within the same cluster), so the resulting term controls for broad regional co-presence rather than for membership in a local scene. Estimates for the spatial overlap term retain their sign and significance for values of $k$ between 6 and 10. Outside this range, the measure degrades in both directions: with fewer clusters, players active in distinct scenes are treated as co-present, while with more, individual clusters contain too few recording years for the overlap proportion to be informative.
 
We model each artist's location history as a discrete time-series. Formally, let $\mathcal{Y} = \{y_{\min}, \dots, y_{\max}\}$ denote the continuous range of years spanned by the dataset. For each artist $a_i$, we construct a trajectory vector $\mathbf{v}_i \in \{0, 1, \dots, K\}^T$, where the element $v_{i,t}$ corresponds to the spatial state of artist $a_i$ in year $t$:
 
\begin{equation*}
    v_{i,t} = 
    \begin{cases} 
    c \in \{1, \dots, 8\} & \text{if artist } a_i \text{ was active in cluster } c \text{ during year } t, \\
    0 & \text{if no activity was recorded for year } t.
    \end{cases}
\end{equation*}
 
To quantify the geographical affinity between artists, we define an asymmetric \textbf{Spatial Overlap Matrix}, $S^{\text{overlap}}$ $\in \mathbb{R}^{N \times N}$. This metric assesses the extent to which the career trajectory of one artist is spatially contained within the trajectory of another. Let $\mathcal{O}_i$ represent the set of active spatiotemporal tuples for artist $a_i$, defined as $\mathcal{O}_i = \{ (t, v_{i,t}) \mid v_{i,t} \neq 0 \}$. The directed overlap $S^{\text{overlap}}_{ij}$ (measuring the overlap of $i$ relative to the career of $j$) is defined as:
 
\begin{equation*}
    S^{\text{overlap}}_{ij} = \frac{|\mathcal{O}_i \cap \mathcal{O}_j|}{|\mathcal{O}_j|}
\end{equation*}
\noindent
 
The numerator $|\mathcal{O}_i \cap \mathcal{O}_j|$ represents the number of years where both artists were active in the same spatial cluster. The denominator $|\mathcal{O}_j|$ is the total count of active years for the receiving artist $a_j$. This asymmetry is critical for analyzing artists with disparate career lengths. It ensures that if a short-career artist $a_j$ performed exclusively in locations shared with a long-career artist $a_i$, the overlap score $S_{ij}$ will be 1 (indicating total containment of $j$'s spatial history within $i$'s), whereas $S^{\text{overlap}}_{ji}$ would be significantly lower.
 
Two features of this construction should be noted. Years in which no session is recorded for an artist contribute to neither the numerator nor the denominator, so the measure is defined over observed activity rather than over the full span of a career. Where an artist is recorded in more than one cluster within a single year, we assign the cluster in which the largest number of that year's sessions took place. We provide a visualization of the spatial activity of jazz recordings in the Appendix; see Figure~\ref{fig:activity_map}.

\subsection{Temporal Overlap}

To separate contemporaneity from co-location, we also measure how far two artists' recording careers overlap in time, irrespective of where they recorded. Let $R_j$ denote the set of recording sessions of artist $a_j$, let $t(s)$ denote the calendar year of session $s$, and let $T_i = \{\, t(s) \mid s \in R_i \,\}$ be the set of years in which artist $a_i$ recorded at least once. We define the asymmetric \textbf{Temporal Overlap Matrix} $S^{\text{time}}] \in [0,1]^{N \times N}$ as
\[
S^{\text{time}}_{ij} = \frac{\lvert \{\, s \in R_j \mid t(s) \in T_i \,\} \rvert}{\lvert R_j \rvert}.
\]
The numerator counts the sessions of artist $a_j$ that took place in a year in which artist $a_i$ was also active, and the denominator is the total number of sessions of $a_j$. $S^{\text{time}}_{ij}$ thus measures the share of $j$'s recorded output that falls within $i$'s active years. As with spatial overlap, the measure is asymmetric: if a short-career artist $a_j$ recorded only during the active years of a long-career artist $a_i$, then $S^{\text{time}}_{ij} = 1$, whereas $S^{\text{time}}_{ji}$ may be considerably lower. Because the denominator counts sessions rather than active years, years in which $a_j$ recorded intensively weigh more than years with a single session. In the models, we use the standardized version $\tilde{S}^{\text{time}}_{ij}$.

\section{Methods}

Exponential random graph models (ERGMs) offer a flexible statistical framework for investigating network structure \citep{lusher_exponential_2013, robins_exponential_2011}. In this study, we leverage ERGMs to examine how different organizing principles jointly structure the relational anatomy of cultural influence among jazz trumpet players. Following past accounts on networks in cultural fields and our theoretical considerations, we assume that influence should be studied as a relational outcome shaped by multiple organizing principles at once: opportunity structures, similarity in cultural content, reputational asymmetries, and the endogenous dynamics of network formation (for example, concentration of attention on already prominent figures). ERGMs can capture this complex interplay because they allow explicit formulation of dependency assumptions in the model specification, rather than treating dyads as independent.

By conceptualizing the overall structure of the observed influence network as the dependent variable, ERGMs can include terms based on multiple local structures as independent variables, yielding parameter estimates that depict various tie-formation mechanisms organizing the structure of cultural influence \citep{desmarais_micro-level_2012}. Intuitively, one can think of local structures as ingredients that jointly constitute the observed network. The estimated parameters indicate whether a particular local structure—for instance, dyads between musicians sharing the same generation—contributes significantly to predicting the network's overall structure, given all other parameters in the model \citep{lusher_formation_2013}. In our case, we include terms that capture endogenous network dynamics, as well as exogenous network mechanisms, such as overlap in shared collaborators and sonic context similarity, that may structure influence relationships among trumpet players. The flexibility of the ERGM framework also allows us to introduce interaction terms and terms that reflect the likelihood of observing an influence tie among players who are highly similar both socially and in sonic context. Thereby, we can assess how social and sonic context similarity jointly structure influence ties and whether observing an influence tie becomes less likely when musicians are highly similar. 

Within the ERGM framework, we treat the influence network as a dependent realization of a global probability distribution. Using Monte Carlo Maximum Likelihood Estimation (MCMLE), we estimate coefficients for structural terms, such as degree distributions and shared partners, where the sampling process is conducted with a toggle algorithm. Model convergence is achieved when the structural properties of the simulated networks align with the observed jazz influence network. This generative approach provides a relationally oriented model for network structures that acknowledges the interdependent nature of network data and accounts for both the opportunity structure for tie formation and network-endogenous mechanisms.

Unlike regression techniques, which assume independence of residuals \citep{angrist_mostly_2008}, ERGMs explicitly address dependencies in network data. This is particularly important for understanding influence networks, where the presence of one influence tie may make others more or less likely. For example, if trumpet players B, C, and D were all influenced by player A,
this may increase the likelihood that a further player E is also influenced
by A; a pattern of preferential attachment reflecting that some trumpet
players accumulate disproportionate influence (e.g., Miles Davis).

\subsection{Average Marginal Effects}
ERGM coefficients are expressed on the scale of the conditional log-odds of a
tie, given the rest of the network. Because the logistic link is nonlinear, the
same coefficient implies different changes in probability for dyads with
different baseline probabilities; and because coefficients in logit-type models
are identified only relative to the unobserved residual variance, their
magnitudes cannot be compared directly across nested specifications
\citep{mood_logistic_2010, duxbury_problem_2023}. We therefore complement the
coefficients with average marginal effects (AMEs)
\citep{duxbury_problem_2023, duxbury_ergmargins_2019}. For each dyad, the
marginal effect of a predictor is the change in the conditional probability of
an influence tie, given the rest of the observed network, associated with a
one-unit change in that predictor; the AME averages this quantity across all
2{,}652 dyads.

All continuous dyadic predictors are $z$-standardized, so their AMEs refer to a
one-standard-deviation change. For binary predictors (direct collaboration and
the top-decile indicators), we report the discrete change from 0 to 1
\citep{long_regression_1997}. Sonic context similarity and collaborator overlap also
enter the model through their product, so the slope of each varies with the
value of the other; their dyad-level marginal effects therefore incorporate the
interaction term. Because the coefficient of a product term does not equal the
interaction effect on the probability scale \citep{ai_interaction_2003}, we
report overall AMEs, conditional AMEs at selected percentiles of the moderator,
and second differences between them \citep{mize_best_2019}. AMEs for the nodal
covariates refer to each term (sender, receiver, absolute difference) with the
other two held fixed. We do not report AMEs for the geometrically weighted degree terms, because a structural statistic cannot be changed while holding the rest of the network constant; \citet{duxbury_marginal_2026} proposes marginal structural effects for such terms, which we do not compute because our research questions concern the dyadic covariates, for which the degree terms serve as controls. Uncertainty is
quantified by drawing 2{,}000 parameter vectors from the estimated sampling
distribution of the coefficients and reporting 95\% percentile intervals
\citep{king_making_2000}.

AMEs serve two purposes here. First, they express associations in percentage
points, which can be set against the baseline tie probability. Second, AMEs are
considerably less sensitive than coefficients to the rescaling that occurs when
predictors are added \citep{mood_logistic_2010, duxbury_problem_2023}, which
makes descriptive comparisons across nested specifications meaningful; we use
this to assess, for instance, whether the association between collaborator
overlap and influence ties changes once sonic context similarity is included.

\subsection{Model specifications}

We build the model additively and iteratively, which lets us address our research questions, manage model complexity, and reduce the risk of degeneracy \citep{wimmer_beyond_2010, lusher_exponential_2013}. At each step, we assess model fit via AIC and goodness-of-fit diagnostics and verify satisfactory MCMC mixing. We begin with a baseline specification including the edges term and geometrically weighted in- and out-degree statistics, capturing network density and heterogeneity in incoming and outgoing ties. We then progressively introduce nodal attributes and edge-level covariates, enabling us to assess their contributions conditional on the underlying structural effects. Based on this procedure, we present three models that differ in their inclusion of nodal attributes and relational similarity measures, offering a focused comparison of collaboration-based, sonic-context-based, and combined mechanisms of tie formation. Formally, we specify the full ERGM as
{\allowdisplaybreaks
\begin{align*}
& P(Y = y \mid X) \propto \exp\Big( \\
&\quad \theta_1 \sum_{i \neq j} y_{ij}
  + \theta_2 \Big[ e^{\alpha_{\text{in}}} \sum_{k=1}^{n-1} \big( 1 - (1 - e^{-\alpha_{\text{in}}})^k \big) D_k^{\text{in}}(y) \Big] \\
&\quad + \theta_3 \Big[ e^{\alpha_{\text{out}}} \sum_{k=1}^{n-1} \big( 1 - (1 - e^{-\alpha_{\text{out}}})^k \big) D_k^{\text{out}}(y) \Big]
  + \theta_4 \sum_{i \neq j} y_{ij} \tilde{r}_i \\
&\quad + \theta_5 \sum_{i \neq j} y_{ij} \tilde{r}_j
  + \theta_6 \sum_{i \neq j} y_{ij} \tilde{b}_i
  + \theta_7 \sum_{i \neq j} y_{ij} \tilde{b}_j
  + \theta_8 \sum_{i \neq j} y_{ij} \tilde{S}^{\text{collab}}_{ij} \\
&\quad + \theta_9 \sum_{i \neq j} y_{ij} S^{\text{collab-direct}}_{ij}
  + \theta_{10} \sum_{i \neq j} y_{ij} \tilde{S}^{\text{overlap}}_{ij}
  + \theta_{11} \sum_{i \neq j} y_{ij} \tilde{S}^{\text{sonic}}_{ij} \\
&\quad + \theta_{12} \sum_{i \neq j} y_{ij} \lvert \tilde{r}_i - \tilde{r}_j \rvert
  + \theta_{13} \sum_{i \neq j} y_{ij} \lvert \tilde{b}_i - \tilde{b}_j \rvert \\
&\quad + \theta_{14} \sum_{i \neq j} y_{ij} \big( \tilde{S}^{\text{sonic}}_{ij} \cdot \tilde{S}^{\text{collab}}_{ij} \big)
  + \theta_{15} \sum_{i \neq j} y_{ij} \,\mathbf{1}\!\left[\tilde{S}^{\text{sonic}}_{ij} > \tau^{\text{sonic}}\right] \\
&\quad + \theta_{16} \sum_{i \neq j} y_{ij} \,\mathbf{1}\!\left[\tilde{S}^{\text{collab}}_{ij} > \tau^{\text{collab}}\right]
  + \theta_{17} \sum_{i \neq j} y_{ij} \tilde{S}^{\text{time}}_{ij}
\Big).
\end{align*}
}
\noindent

Here, $P(Y = y \mid X)$ denotes the probability of observing the realized influence network $y$ from the ensemble of all possible directed networks on the same node set, conditional on covariates $X$. The term $y_{ij}$ indicates the presence of a directed influence tie from artist $i$ to artist $j$. The matrices $\tilde{S}^{\text{collab}}_{ij}$, $S^{\text{collab-direct}}_{ij}$, $\tilde{S}^{\text{sonic}}_{ij}$, $\tilde{S}^{\text{overlap}}_{ij}$, and $\tilde{S}^{\text{time}}_{ij}$ represent, respectively, the standardized share of shared collaborators, the direct collaboration matrix between artists, the standardized sonic context similarity (the rescaled cosine similarity between artists' mean 128-dimensional audio embeddings), the standardized spatial activity overlap, and the standardized temporal overlap between artists $i$ and $j$. All edge-level covariates except direct collaboration are standardized to unit variance. The terms $\tilde{S}^{\text{collab}}_{ij}$ and $\tilde{S}^{\text{sonic}}_{ij}$ are of particular interest since they reflect the extent to which ascribed influence intersects with having overlapping circles of collaborators and recording in similar sonic contexts. Including them simultaneously allows us to address RQ1, which asks whether an association between ascribed influence and social similarity holds after accounting for similarity in musicians' sonic context. The terms $\tilde{S}^{\text{overlap}}_{ij}$ and $\tilde{S}^{\text{time}}_{ij}$ serve as controls: the former reflects whether musicians who recorded in the same regions during the same years are more likely to share an influence tie, and the latter whether musicians whose recording careers overlap in time, irrespective of place, are more likely to do so.

At the nodal level, $\tilde{r}_i$ denotes the total number of releases of artist $i$, standardized to unit variance, while $\tilde{b}_i$ denotes the birth year of artist $i$, also standardized to unit variance. The corresponding vectors $\mathbf{r}$ and $\mathbf{b}$ lie in $\mathbb{R}^{1 \times N}$, where $N = 52$, the number of artists in the network. Absolute difference terms $\lvert \tilde{r}_i - \tilde{r}_j \rvert$ and $\lvert \tilde{b}_i - \tilde{b}_j \rvert$ capture assortative or disassortative mixing effects with respect to productivity and cohort similarity. These terms control for the facts that successful artists are more influential, and that influence tends to flow between rather than within generations of musicians. 

To examine whether the association between similarity and ascribed influence weakens at the upper end of each distribution (RQ2), we include two further indicator terms. The first, $[\tilde{S}^{\text{sonic}}_{ij} > \tau^{\text{sonic}}]$, flags dyads whose standardized sonic context similarity exceeds a cutoff $\tau^{\text{sonic}}$, allowing us to test whether very high similarity in sonic context is associated with a lower rather than higher probability of an ascribed relation. The second, $[\tilde{S}^{\text{collab}}_{ij} > \tau^{\text{collab}}]$, analogously flags dyads with unusually dense shared collaboration networks beyond a cutoff $\tau^{\text{collab}}$. Both cutoffs are defined on the standardized scale. We evaluate a range of values for each threshold parameter; full results across all cutoff specifications are reported in the Appendix.

Finally, to address RQ3, we include an interaction term between $\tilde{S}^{\text{collab}}_{ij}$ and $\tilde{S}^{\text{sonic}}_{ij}$, which tests whether the probability of an ascribed relation between two players who both share collaborators and have similar sonic contexts is lower than either condition alone would imply. Because we control for spatio-temporal overlap ($\tilde{S}^{\text{overlap}}_{ij}$), the interaction does not merely pick up players who were active in the same places at the same time. We defer the interpretation of this term to the Discussion, where we consider the behavioral and classificatory readings it admits.

The three reported specifications correspond to nested versions of this model. In the collaboration-only model, we exclude the sonic context similarity term $\tilde{S}^{\text{sonic}}_{ij}$, its interaction, and the sonic context threshold indicator. In the sonic-context-only model, we instead exclude the collaboration overlap term $\tilde{S}^{\text{collab}}_{ij}$, direct collaboration term $S^{\text{collab-direct}}_{ij}$ and the collaboration threshold indicator, reparameterizing the model accordingly. The full model includes both similarity measures, their interaction, and both threshold indicators, allowing us to assess how sonic context similarity and shared collaboration jointly structure influence ties, and whether extremely high similarity has a suppressive effect consistent with the idea that a balance of similarity and distance is necessary for influence to flow: While collaborating with similar musicians and recording in similar sonic contexts establishes a common thread in the tradition, occupying the same coordinates in social space and in sonic context should mitigate the likelihood of influence, because it indicates that musicians are imitators rather than innovators who honor tradition but also advance it. 

\section{Results}
\subsection{Descriptive structure of the genealogy}
\label{sec:descriptive}

\begin{figure}[h!]
    \centering
    \includegraphics[width=0.99\linewidth]{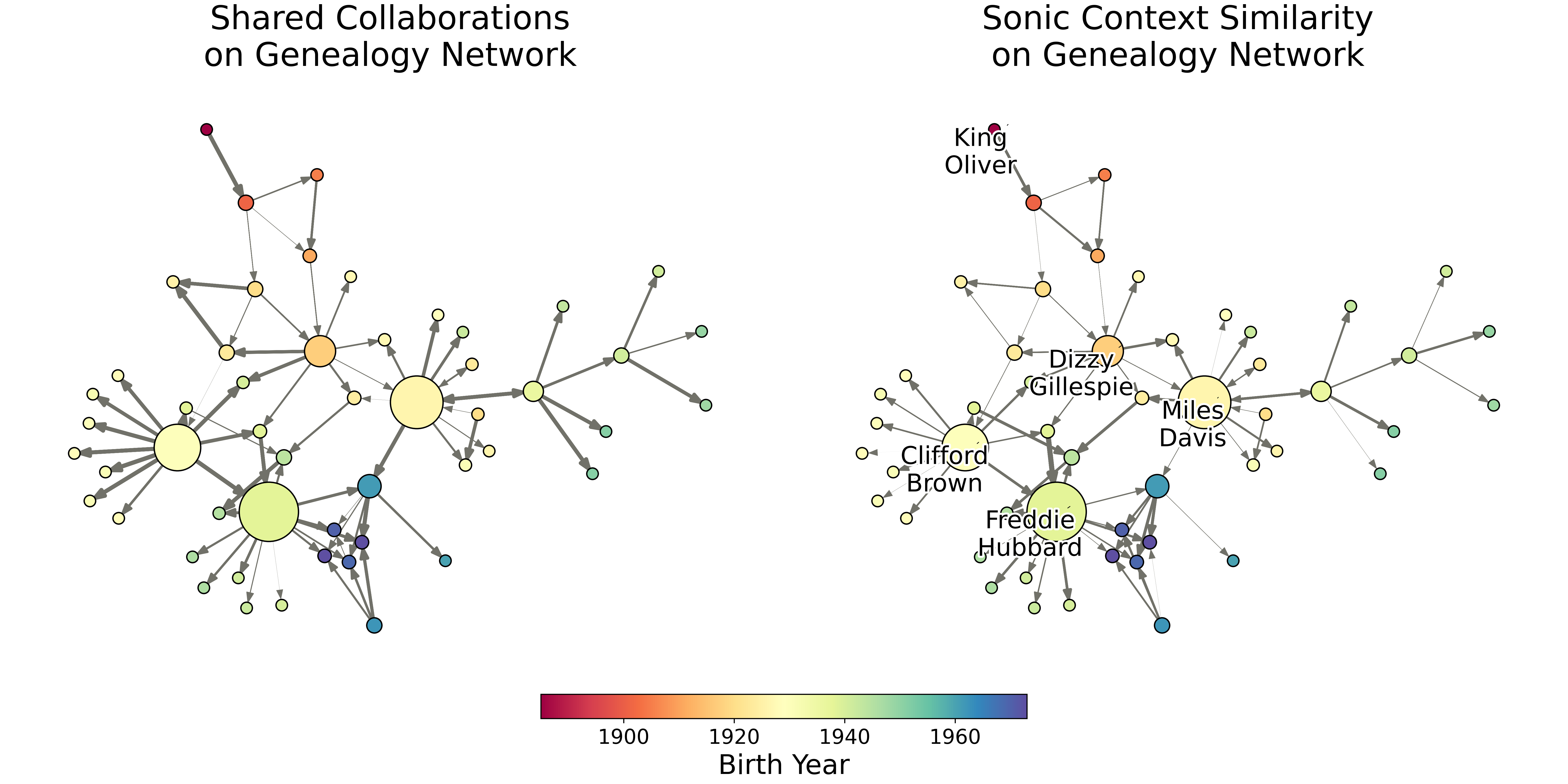}
\caption{\textbf{Comparing genealogical influence with collaboration- and sonic-context-based proximity.} Both networks demonstrate the directed influence relationship between trumpet players. The left network reweights the edges using collaboration overlap, with thicker edges indicating a greater number of shared collaborators. The right panel reweights edges by sonic context similarity, the similarity of the recordings on which two players appear, computed from VGGish embeddings of the full recordings.}
\label{fig:collab_gen_audio}
\end{figure}

The first step in our analysis is to map how strongly influence among genealogy players overlaps with shared collaboration partners and similarity in sonic context (Figure~\ref{fig:collab_gen_audio}). The left panel of Figure~\ref{fig:collab_gen_audio} depicts the influence network and adjusts the weights of ties according to how many collaborators two trumpet players shared in their career. For instance, the players influenced by Clifford Brown all show a high overlap in collaborators with him, indicating that they were active in the same social circles. The right panel of Figure~\ref{fig:collab_gen_audio} shows the same influence network, but here edge weight reflects similarity in players' sonic profiles. Comparing both panels indicates that while some influence ties overlap with sonic context similarity, other parts of the network correspond more closely to proximity in social space, measured by recording with a similar set of musicians. This observation is in line with the notion that proximity in social space and in sonic context can both contribute to understanding the structure of recognized influence ties, and we will assess this aspect more systematically in our ERGM analysis.

\begin{figure}[H]
    \centering
    \includegraphics[width=0.99\linewidth]{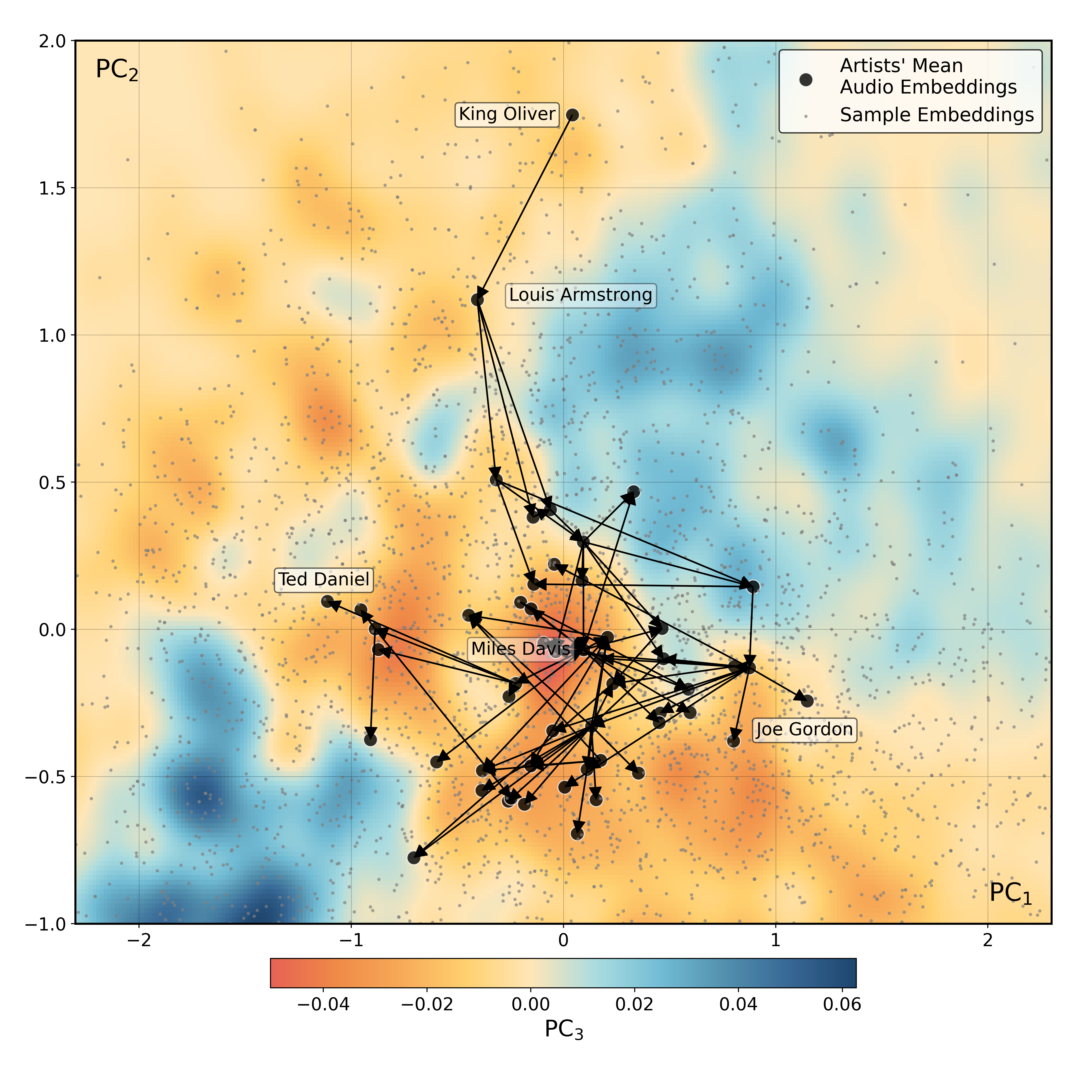}
\caption{\textbf{Descriptive projection of the audio embeddings.} Artist nodes are positioned by their sonic context profiles (mean VGGish embeddings), projected onto the first three principal components (55\% of the embedding variance), and connected by the genealogy network. The projection is illustrative: distances in the figure are not distances in the full 128-dimensional space, on which all statistical analyses are based. The components appear to reflect differences in timbre, recording era, and style, separating, for example, early acoustic recordings (King Oliver, 1920s) from experimental palettes (Ted Daniel, 1970s) and modern studio tones (Joe Gordon, 1950s--60s).}
\label{fig:PCA}
\end{figure}

We now turn to the relation between recognized influence and sonic context similarity in greater detail. Figure~\ref{fig:PCA} shows a lower-dimensional representation of the VGGish audio embeddings and indicates the positions of genealogy players, as well as the influence ties between them. Specifically, we project the 128-dimensional audio embeddings onto the first three components of a principal component analysis (PCA). Subsequently, we compute average coordinates for each trumpet player in this reduced space by averaging over the vectors attached to all of their recordings and overlay the genealogy network of documented influence ties. Note that the visualization only captures part of the variance of the embedding space ($\approx55\%$).
\noindent

The figure helps us see how recognized influence relates to closeness and distance in the sonic characteristics of players' recordings. Interestingly, the visualization shows that most influence ties span a certain distance rather than clustering tightly. For example, Louis Armstrong and his mentor King Oliver both occupy the middle-top area of the first two dimensions, but with a visible shift in coordinates from Oliver to Armstrong. Similar shifts appear across subsequent generations, which indicates that recognized influence entails a balance between similarity and differentiation: while the overall shape of McNeil's genealogy is preserved when mapped onto the sonic context space, players who furthered the lineage of jazz trumpet playing are separated by jumps, whereas the musicians they influenced tend to appear closer to one another.   

Furthermore, upon closer inspection \footnote{We manually examined a selection of recordings from extreme regions of each PCA dimension to qualitatively grasp which characteristics structure the embedding space. While this approach is sufficient for our descriptive purposes here, further research could pursue a more systematic approach to quantify the regions from our embeddings dataset that can be found in the repository.}, we find that the low end of $\text{PC}_1$ holds vocal-led tracks, ballads, and speech (Sinatra, Astaire, Armstrong vocals). The high end holds bebop and hard-bop small combos (Parker, Mobley, Navarro, Blakey). The high end of $\text{PC}_2$ is populated by 1920s–30s recordings (King Oliver, early Armstrong); the low end seems to capture post-1975 production (fusion, smooth jazz, remixes). Consequently, we suspect that the space partially reflects recording techniques---for instance, decisions in microphone placement, mixing, and mastering---rather than the session date. Ensemble size also seems to organize the space. The high end of $\text{PC}_3$ is marked by big-band and orchestral recordings (Herman, Kenton, Basie, film scores). Part of $\text{PC}_3$ also reflects loudness (it tracks Deezer's gain field) and the curvature of the point cloud. It is the least clean of the three axes. To move beyond this descriptive view, the next sections model the relationship between recognized influence, collaborations, and sonic context similarity using statistical network models.

\subsection{Do social proximity and sonic context similarity each contribute to explaining recognized influence?}
While the analysis presented so far focused on the interplay between recognized influence, shared collaborators, and sonic context similarity, we next turn to statistical network models to quantify which aspects explain the relational structure of influence. Broadly speaking, we find that social proximity \textit{and} sonic context similarity each contribute to explaining ascribed influence relations (Table~\ref{tab:ergm_models}).\footnote{Nested comparisons confirm that neither set of predictors is redundant. Adding the collaboration terms to the sonic context model reduces the deviance by 26.7 on 4 degrees of freedom, and adding the sonic context terms to the collaboration model reduces it by 35.5 on 3 degrees of freedom (both $p < .001$; approximate, given MCMC-MLE estimation). The full model also has the lowest AIC (389.6, against 408.3 for the sonic context and 419.1 for the collaboration model), although BIC, which penalizes additional parameters more heavily, slightly favors the sonic context model (484.8 against 489.6).} Regarding social proximity, we find that direct collaboration intersects strongly with recognized influence ties. Descriptively, 46\% of ascribed relations join players who recorded together, compared with 16\% of all dyads, which indicates that direct collaboration is more frequent among influence dyads. Our ERGM analyses are in line with this descriptive finding. In the full model, having recorded together is associated with 2.8 times higher conditional odds of an ascribed influence tie. Note that this holds net of collaborator overlap, which is our second measurement for social proximity. We find that collaborator overlap is also positively associated with recognized influence net of direct collaboration, both in the collaboration model (0.45, SE 0.21) and in the full model (0.72, SE 0.27).\footnote{Because the full model includes the interaction between sonic context similarity and collaborator overlap, the main-effect coefficients of these two measures refer to conditional associations at the mean of the other, $z$-standardized measure: at average collaborator overlap, the coefficient for sonic context similarity is 3.81 (SE 0.85), and at average sonic context similarity, the coefficient for collaborator overlap is 0.72 (SE 0.27).} Taken together, these findings partially address our first research question and highlight that social proximity contributes to explaining recognized influence above and beyond other factors, such as musicians' sonic profiles and network endogenous mechanisms.

\begin{table}[htbp]
\centering
\providecommand{\ci}[2]{\multicolumn{1}{c}{\scriptsize[$#1$,\,$#2$]}}
\begin{threeparttable}
\caption{ERGM Model Results: Hierarchical Comparison}
\label{tab:ergm_models}
\footnotesize
\setlength{\tabcolsep}{4pt}
\begin{tabular*}{\ergmTableWidth}{@{\extracolsep{\fill}}lCACACA}
\toprule
 & \multicolumn{2}{c}{Sonic Context Model} & \multicolumn{2}{c}{Collab Model} & \multicolumn{2}{c}{Full Model} \\
\cmidrule(lr){2-3} \cmidrule(lr){4-5} \cmidrule(lr){6-7}
 & \multicolumn{1}{c}{Coef.} & \multicolumn{1}{c}{AME} & \multicolumn{1}{c}{Coef.} & \multicolumn{1}{c}{AME} & \multicolumn{1}{c}{Coef.} & \multicolumn{1}{c}{AME} \\
\midrule
Edges & -2.31^{***} & \multicolumn{1}{c}{---} & -1.62^{***} & \multicolumn{1}{c}{---} & -3.26^{***} & \multicolumn{1}{c}{---} \\
 & (0.45) & & (0.36) & & (0.58) & \\
GWID ($\alpha=0.1$) & 3.30^{***} & \multicolumn{1}{c}{---} & 3.72^{***} & \multicolumn{1}{c}{---} & 3.28^{***} & \multicolumn{1}{c}{---} \\
 & (0.87) & & (0.91) & & (0.86) & \\
GWOD ($\alpha=0.3$) & -3.83^{***} & \multicolumn{1}{c}{---} & -3.96^{***} & \multicolumn{1}{c}{---} & -3.83^{***} & \multicolumn{1}{c}{---} \\
 & (0.50) & & (0.49) & & (0.51) & \\
\midrule
\multicolumn{7}{l}{\textit{Edge covariates}} \\
Overlap time & -0.55 & -0.8 & -0.92 & -1.4 & -0.94 & -1.2 \\
 & (0.60) & \ci{-2.5}{1.0} & (0.62) & \ci{-3.3}{0.4} & (0.65) & \ci{-3.0}{0.5} \\
Overlap space & -0.14 & -0.2 & -0.54^{***} & -0.8 & -0.41^{*} & -0.5 \\
 & (0.11) & \ci{-0.6}{0.1} & (0.16) & \ci{-1.4}{-0.4} & (0.17) & \ci{-1.0}{-0.1} \\
Sonic context similarity$^{b}$ & 2.57^{***} & 3.7 & & & 3.81^{***} & 3.7 \\
 & (0.55) & \ci{2.1}{5.3} & & & (0.85) & \ci{2.1}{5.7} \\
Sonic context (top)$^{a}$ & -0.78^{\dagger} & -1.0 & & & -0.78^{\dagger} & -0.9 \\
 & (0.46) & \ci{-1.9}{0.2} & & & (0.47) & \ci{-1.8}{0.2} \\
Direct collab.$^{a}$ & & & 0.80^{*} & 1.3 & 1.02^{**} & 1.5 \\
 & & & (0.34) & \ci{0.3}{3.0} & (0.35) & \ci{0.5}{2.9} \\
Shared collab.$^{b}$ & & & 0.45^{*} & 0.7 & 0.72^{**} & 0.1 \\
 & & & (0.21) & \ci{0.1}{1.3} & (0.27) & \ci{-0.5}{0.7} \\
Collab (top)$^{a}$ & & & 0.10 & 0.1 & 0.29 & 0.4 \\
 & & & (0.44) & \ci{-1.0}{1.8} & (0.47) & \ci{-0.8}{2.0} \\
Sonic context $\times$ Shared collab. & & & & & -1.14^{**} & \multicolumn{1}{c}{---$^{c}$} \\
 & & & & & (0.35) & \\
\midrule
\multicolumn{7}{l}{\textit{Node covariates: releases}} \\
Sender & 0.72^{**} & 1.0 & 0.39^{\dagger} & 0.6 & 0.77^{**} & 1.0 \\
 & (0.25) & \ci{0.3}{1.8} & (0.23) & \ci{-0.1}{1.2} & (0.27) & \ci{0.3}{1.8} \\
Receiver & 1.44^{***} & 2.0 & 1.54^{***} & 2.3 & 1.35^{***} & 1.7 \\
 & (0.33) & \ci{1.2}{3.0} & (0.32) & \ci{1.4}{3.3} & (0.35) & \ci{0.9}{2.7} \\
Abs.\ difference & -0.33 & -0.5 & -0.14 & -0.2 & -0.41 & -0.5 \\
 & (0.29) & \ci{-1.3}{0.4} & (0.27) & \ci{-1.0}{0.7} & (0.31) & \ci{-1.4}{0.3} \\
\midrule
\multicolumn{7}{l}{\textit{Node covariates: year of birth}} \\
Sender & -3.71^{***} & -5.3 & -3.64^{***} & -5.4 & -3.59^{***} & -4.6 \\
 & (1.05) & \ci{-8.5}{-2.6} & (0.94) & \ci{-8.6}{-2.9} & (1.04) & \ci{-7.7}{-2.2} \\
Receiver & 5.43^{***} & 7.7 & 5.21^{***} & 7.7 & 5.12^{***} & 6.6 \\
 & (1.08) & \ci{5.0}{11.0} & (0.99) & \ci{5.0}{11.1} & (1.07) & \ci{4.2}{10.0} \\
Abs.\ difference & -5.65^{***} & -8.0 & -5.50^{***} & -8.2 & -5.32^{***} & -6.9 \\
 & (1.10) & \ci{-11.6}{-5.2} & (1.00) & \ci{-11.7}{-5.5} & (1.10) & \ci{-10.3}{-4.2} \\
\midrule
Dyads & \multicolumn{2}{c}{$2{,}652$} & \multicolumn{2}{c}{$2{,}652$} & \multicolumn{2}{c}{$2{,}652$} \\
Residual deviance & \multicolumn{2}{c}{$382.3$} & \multicolumn{2}{c}{$391.1$} & \multicolumn{2}{c}{$355.6$} \\
AIC & \multicolumn{2}{c}{$408.3$} & \multicolumn{2}{c}{$419.1$} & \multicolumn{2}{c}{$389.6$} \\
BIC & \multicolumn{2}{c}{$484.8$} & \multicolumn{2}{c}{$501.5$} & \multicolumn{2}{c}{$489.6$} \\
\bottomrule
\end{tabular*}
\begin{tablenotes}
\footnotesize
\item Note: Coefficients in log-odds; standard errors in parentheses; *** p$<$0.001, ** p$<$0.01, * p$<$0.05, $^\dagger$ p$<$0.1. AME: average change across dyads in the conditional probability of a tie (percentage points) per one-SD increase in the predictor (continuous predictors $z$-standardized); 95\% simulation intervals in brackets (2{,}000 draws). Node-covariate AMEs hold the other two terms fixed. $^{a}$Discrete change 0$\to$1. $^{b}$Full model: overall AME including the interaction. $^{c}$No separate marginal effect; for $^{b}$ and $^{c}$ see Table~\ref{tab:interaction}. AMEs are not reported for edges, GWID, and GWOD (see Section~5.1).
\end{tablenotes}
\end{threeparttable}
\end{table}

\begin{table}[htbp]
\centering
\providecommand{\ci}[2]{\multicolumn{1}{c}{\scriptsize[$#1$,\,$#2$]}}
\begin{threeparttable}
\caption{Average Marginal Effects of Sonic Context Similarity by Collaborator Overlap (Full Model)}
\label{tab:interaction}
\footnotesize
\begin{tabular*}{\ergmTableWidth}{@{\extracolsep{\fill}}lCACACA}
\toprule
 & AME (pp) & 95\% interval \\
\midrule
\multicolumn{3}{l}{\textit{Conditional AMEs of sonic context similarity}} \\
Collaborator overlap at P10 ($-0.93$ SD) & 5.99 & [3.19, 10.59] \\
Collaborator overlap at P50 ($-0.30$ SD) & 5.31 & [2.98, 8.58] \\
Collaborator overlap at P90 ($1.39$ SD) & 3.33 & [1.65, 5.42] \\
Collaborator overlap at P99 ($3.20$ SD) & 0.38 & [$-4.95$, 3.49] \\
\midrule
\multicolumn{3}{l}{\textit{Second differences}} \\
P50 vs.\ P10 & $-0.69$ & [$-2.18$, 0.06] \\
P90 vs.\ P10 & $-2.67$ & [$-6.80$, $-0.04$] \\
P99 vs.\ P10 & $-5.61$ & [$-12.66$, $-1.05$] \\
\bottomrule
\end{tabular*}
\begin{tablenotes}
\footnotesize
\item Note: Conditional AMEs fix collaborator overlap at the indicated percentile for all dyads and average the marginal effect of a one-standard-deviation increase in sonic context similarity, including the interaction term. Second differences are differences between conditional AMEs \citep{mize_best_2019}. 95\% percentile intervals from 2{,}000 draws from the estimated sampling distribution of the coefficients.
\end{tablenotes}
\end{threeparttable}
\end{table}

Regarding sonic context similarity, we find that a recognized influence tie is substantially more likely if the recordings in which two trumpet players participated sound similar. In particular, a one-standard-deviation increase in sonic context similarity is associated with a 3.7 percentage-point higher conditional probability of an ascribed influence tie---which is a considerable effect size given that the baseline probability of a tie is 2.8 percent. Interestingly, the average marginal effect for context similarity retains its size in the full model, which further corroborates the notion that social and sonic context proximity contribute independently to explaining the structure of recognized influence.\footnote{The AME of direct collaboration is similarly stable (1.3 percentage points in the collaboration model; 95\% CI: 0.3--3.0). By contrast, the AME of collaborator overlap falls from 0.7 percentage points in the collaboration model (95\% CI: 0.1--1.3) to 0.1 points in the full model (95\% CI: $-0.5$ to 0.7). This does not imply that collaborator overlap is irrelevant: its log-odds slope, $0.72 - 1.14 \times$ sonic context similarity, is positive for pairs below about 0.6 standard deviations of sonic context similarity and negative above, and these opposing associations largely cancel when averaged across dyads (Section~\ref{sec:interaction}).}

\subsection{Are highly similar players less likely to be recognized as influences?}
Next, we turn to results addressing our second research question whether extreme levels of social and sonic context similarity are associated with a decrease in the likelihood of an influence tie. We find that dyads in the top decile of sonic context similarity have somewhat lower conditional log-odds of an ascribed influence tie. While the estimate does not reach conventional significance, altering the thresholds from the top decile slightly reveals that the likelihood of an influence tie is significantly lower for the upper 8,9,11, and 12\% of highly similar dyads, see~\ref{app:thresholds}. In contrast, the corresponding indicator for collaborator overlap is insignificant across thresholds. Taken together, this provides tentative evidence that the players with the most similar sonic context are less likely to be linked than their similarity would otherwise predict.

\subsection{What if sonic context and social proximity intersect?}
\label{sec:interaction}
Our third research question asked whether an intersection between social and sonic context similarity decreases the likelihood of a recognized influence tie. Such a pattern would indicate that players are part of the same scene rather than being two exponents in a lineage. We find that the interaction between sonic context similarity and collaborator overlap is negative while both main effects are positive. On the log-odds scale, the association is therefore sub-additive: the slope of sonic context similarity, $3.81 - 1.14 \times$ collaborator overlap, declines as collaborator overlap increases and reaches zero about 3.3 standard deviations above its mean. On the probability scale, the AME of sonic context similarity declines from 6.0 percentage points (95\% CI: 3.2--10.6) at the 10th percentile of collaborator overlap to 3.3 points (95\% CI: 1.7--5.4) at the 90th percentile, and is indistinguishable from zero at the 99th percentile (0.4 points; 95\% CI: $-5.0$ to 3.5; Table~\ref{tab:interaction}).\footnote{The decline between the 10th and 90th percentiles is distinguishable from zero, although only narrowly ($-2.7$ points; 95\% CI: $-6.8$ to $-0.04$), whereas the difference between the 10th percentile and the median is not ($-0.7$ points; 95\% CI: $-2.2$ to 0.1). The moderation is thus clearly supported on the log-odds scale and more weakly on the probability scale, where the nonlinearity of the link partly offsets the declining slope \citep{ai_interaction_2003, mize_best_2019}. The interaction retains its sign in each of the 20 leave-one-player-out refits and in the MRQAP specification (Section~\ref{sec:robustness}).} Taken together, these results indicate that the likelihood for the ascription of influence, while still positive overall, decreases for musicians who are similar both in their social position and sonic context.

\subsection{Controls}

Generational position is strongly associated with the structure of ascriptions. Older players are more likely to be sources of ascribed influence and younger players are more likely to be recipients, and relations are more likely between players close in age, a pattern of succession between adjacent generations. Success, measured by the number of releases, is associated with both sending (0.77, $p < .01$) and receiving (1.35, $p < .001$) relations, while the absolute difference in success is not ($-0.41$). Overlap in active years is not associated with ascription ($-0.94$, n.s.). Overlap in recording locations is not associated with ascription in the sonic context model (AME $-0.2$ percentage points; 95\% CI: $-0.6$ to 0.1) but is negatively associated once the collaboration terms are included ($-0.8$ points in the collaboration model, 95\% CI: $-1.4$ to $-0.4$; $-0.5$ points in the full model, 95\% CI: $-1.0$ to $-0.1$). Because this change appears on the probability scale, it is not an artifact of coefficient rescaling across nested models. The pattern is consistent with suppression: conditional on collaborator overlap, pairs who recorded in the same locations are less likely to be linked by an ascribed influence relation.

\subsection{Structural terms}

The negative out-degree term (GWOD, $\alpha = 0.3$; $-3.83$) lowers the conditional log-odds of a player's first outgoing tie by 3.83 but of a second by only 0.99, consistent with the observed distribution in which most players send no ties and a few send many. The positive in-degree term (GWID, $\alpha = 0.1$; 3.28) raises the conditional log-odds of a player's first incoming tie by 3.28 but of a second by only 0.31, consistent with most players having a single predecessor (both $p < .001$). As explained in Section~5.1, we do not report average marginal effects for these terms.

A shared-predecessor term, which would capture common mentorship, is not distinguishable from zero and worsens fit (~\ref{app:specs}); given how few shared predecessors the genealogy contains, this does not rule out a shared-mentor explanation.

\subsection{Robustness and model fit}
\label{sec:robustness}
The focal associations are robust to the modeling framework and to the removal of individual players. A multiple regression quadratic assignment procedure with Dekker's double semi-partialling (500 permutations), which conditions on the observed network rather than modeling its dependence structure, reproduces the sign of every focal term and the significance of sonic context similarity ($p = .002$), direct collaboration ($p = .010$), collaborator overlap ($p = .024$), and the interaction ($p = .028$;~\ref{app:qap}, Table~\ref{tab:qap}). The sonic context top-decile indicator is significant in the MRQAP ($p = .026$) but not in the ERGM ($p = .09$), and the collaboration top-decile indicator is null in both ($p = .738$ in the MRQAP). Refitting the full model while omitting, in turn, each of the ten players with the highest out-degree, including Hubbard, Brown, Davis, and Gillespie, and ten randomly chosen players leaves every focal estimate with its sign: sonic context similarity ranges from 3.35 to 4.45, direct collaboration from 0.95 to 1.30, collaborator overlap from 0.60 to 1.00, and the interaction from $-1.45$ to $-0.95$ (~\ref{app:jackknife}). Artist-level sonic context profiles are reproducible across random halves of the available audio (median $r = .73$, or $.85$ after Spearman--Brown correction; ~\ref{app:splithalf}).

Goodness-of-fit diagnostics based on 500 simulated networks (~\ref{app:gof}) show that the full model reproduces the in- and out-degree distributions, edgewise shared partners, and the directed triad census, with one modest exception: players with exactly one outgoing tie are more numerous than predicted. The model under-produces long multi-generational chains, with observed counts at geodesic distances of three to five exceeding the simulated means, although no single deviation is statistically significant.

\section{Discussion}

In this study, we decomposed the relational anatomy of ascribed influence among musicians by taking an expert-curated genealogy of jazz trumpet players as a starting point and analyzing a combination of large-scale archival and audio data. Expert-ascribed influence turns out to be jointly structured by direct collaboration, overlap in collaborators, sonic context similarity, network-endogenous processes, and sociodemographic attributes. Each of the three proximity measures is associated with ascription net of the others, so neither social proximity nor proximity in sonic context is a proxy for the other (\textbf{RQ1}). Direct collaboration deserves particular emphasis: nearly half of the recorded relations join players who actually appeared in a recording session together, against a far lower base rate among all dyads, and the term remains significant once collaborator overlap is included. Additionally, co-presence and a common position in the field are distinct conditions, and both surface in expert judgment.

We find that these associations are conditional (RQ3): where sonic context similarity and collaborator overlap are both high, the probability of an ascribed relation is reduced. We interpret this as a distinction between two relational forms that a canon of lineage keeps apart. Musicians who collaborate with the same people and produce recordings with similar sonic contexts occupy the same niche within the broader jazz community: they are peers, competitors, or co-members of a project, and such a configuration leaves no room for the asymmetry a lineage relation requires. A lineage is a different relation---diachronic rather than synchronic, ordered rather than clustered, and molded by difference as much as by continuity. 

Two qualifications attach to this interpretation. A negative interaction coefficient does not by itself establish that dyads high on both dimensions carry a lower absolute probability of ascription than all other dyads. Consequently, we also examined thresholds in similarity for both dimensions separately (\textbf{RQ2}). Here, a negative estimate appears only for sonic context similarity, and the corresponding term for collaborator overlap is insignificant throughout. The negative estimate for the top decile of sonic context similarity is one of forty specifications we examined and does not reach conventional significance at every cutoff. Hence, while we find tentative evidence that recognized influence among musicians with very similar sonic contexts is less likely, a similar sonic context combined with the same circle of collaborators shows a clearer weak association with ascribed influence. 

Why should the joint occurrence of social proximity and sonic context similarity suppress the ascription of influence? The first reason could be behavioral: players whose recordings are nearly indistinguishable in sonic context from those of their idols have an incentive to differentiate in order to establish a sound of their own, since successful disciples go beyond their influences rather than reproducing them. Under this reading, the line between copying and discipleship is a social fact about how musicians act. The second reason concerns how influence is ascribed, which can operate in two ways. Experts may work with an implicit theory of influence that excludes very close resemblance, reserving an influence tie for musicians who have moved the tradition somewhere new. Or they may leave out players whose contribution is already represented by others. Close followers then are absent from the genealogy construction, and with them also the master--follower pairs with high proximity. Since our models take the node set as given, the patterns we observe appear among the players who remain. Consequently, future research could either contrast multiple lineages or compare included vs. excluded players, which is beyond the scope of our article to assess the generalizability of our findings. 

These readings presuppose a view of the dependent relation that our examination of the source supports: a genealogy of this kind is an expert's representation of the field rather than a complete record of influence. Comparing the recorded relations with the interview and musicological material in \citet{berliner_thinking_2009}, we find that some are independently attested there --- Doc Cheatham ``subsequently adopted Louis Armstrong as his principal mentor'' (p.~123); Charlie Shavers taught Dizzy Gillespie the solos he had memorized from Roy Eldridge recordings (p.~96); Lee Morgan ``absorbed and personalized'' Clifford Brown's tonguing technique (p.~128)---while the book also documents relations the figure omits, including one described by McNeil himself. The representation is also revisable: a later edition of \emph{The Art of Jazz Trumpet} carries a chart that drops three players from the pre-war period and adds four more recent ones \citep{mcneil_art_1999}. Yet the ascription is far from arbitrary. The two lines of descent from Eldridge and Shavers, and the players McNeil marks as most influential, are largely the same in both versions: the chart is revised for very early and contemporary players but stable in between, as one would expect of a working account of a field's history. On a related note, our jackknife robustness checks show that if we remove influential players and re-estimate our models, most coefficients remain very similar. Hence, our findings seem to reflect the overall structure of recognized influence rather than idiosyncrasies induced by individual players. 

Taken together, our results are of interest to a stream of research that goes beyond the micro-level outcomes of collaborative creativity---the product of a particular team and the success of the creators involved---and investigates the nature and evolution of entire creative fields \citep{anheier1991acknowledgment, perczel2025careers, wittek_networks_2025}. By studying ascribed influence, we extend previous accounts that examine the structure of collaboration networks to understand how cultural fields are organized. \citet{wittek_networks_2025} shows that artistic status, captured through cinematic references, was more important for network formation among filmmakers during the New Hollywood era than in other historical periods, and \citet{perczel2025careers} finds that networks of galleries shape pathways of global artistic careers and the logic of consecration in the global field of fine arts. 

Our study also has implications for work that aims to connect field-theoretical perspectives on cultural production with network theory and network-analytic techniques \citep{bottero_worlds_2011, borkenhagen_status_2018}. Where field theory tends toward positional analysis, treating fields as crystallized arenas of competition, the relation we model is vertical: it links positions across generations rather than within one. We are careful about how far this warrants claims about field dynamics. Our design is retrospective and cross-sectional, and our predictors aggregate substantial portions of musicians' careers, so we do not observe a field changing.

In addition, comparing our models yields a practical implication. Overlap in collaboration networks recovers much of the structure of the ascribed influence network even when sonic context information is absent from the specification. We do not read this as showing that social structure produces influence; the two carry distinct information, as the full model shows. But it suggests that when audio is unavailable, the social structure of a cultural field can serve as a partial proxy for its content. This matters for empirical work on music, since streaming platforms have become markedly more restrictive about sharing audio and audio-derived features with researchers.

\subsection{Limitations}

While our article offers new insights into the relational structure of cultural influence, it has several limitations. First, we did not seek to identify causal mechanisms. We focused on dissecting the relational anatomy of ascription rather than isolating factors that make individual artists more or less influential, for two reasons. The genealogy is a non-random selection of highly successful artists from the population of jazz trumpet players, and several of the variables involved, such as success and generational status, are endogenous to its construction. And the decomposition is an achievement in its own right: it shows that multiple organizing principles jointly predict the ascription of influence by jazz experts, revealing features of the relation that historical and qualitative accounts do not make explicit. Our study is therefore complementary to work asking about the causal effects of sonic context similarity or shared collaborators. Future research mobilizing larger datasets with finer temporal resolution, or leveraging natural experiments, could pursue designs that support such claims; integrating a counterfactual notion of causality with statistical network models remains an open methodological problem \citep{an2022causal}.

Second, this is a case study of ascribed influence among jazz musicians, and specifically among trumpet players, so how far our findings extend to other instruments, to jazz more broadly, or to other cultural fields is an open question. We chose this case not because the genealogy is exhaustive but because it is unusually well documented: its author set out his criteria in print, and an independent ethnographic record of the same community exists against which it can be read. Its boundary is nonetheless a canonical selection with no stated population, and it therefore excludes influence from musicians on other instruments as well as from composers, arrangers, producers and bandleaders. The relations it records also carry more information than the relations it omits: McNeil states that Gillespie's influence was too pervasive to draw, and that players pictured with few influences are not thereby without them \citep[25]{mcneil_art_1999}. 

A related limitation concerns how the dependent variable is constituted. Expert ascription is arguably less exposed than self-report to a musician's own incentive to claim or disown an artistic ancestor, but it is retrospective: a tie is assigned once a musician's reputation and place in the canon have settled, so the measure is sensitive to declaration and posture as much as to whatever transmission took place between two players.

A canon's boundary also implicitly entails a field's opportunity structure, shaping who can enter a lineage. For example, our node set contains no female trumpet players. McNeil addresses the absence directly, noting that at the time of writing only Ingrid Jensen ``can be termed truly influential'' among women trumpeters, and attributing the pattern to ``the same conditions (and conditioning) that resulted in women's absence from other professions,'' naming harassment, discriminatory hiring, and the lack of role models within the jazz community \citep[26]{mcneil_art_1999}. Gendered inequality in the production and consumption of jazz is well documented \citep{mcandrew_man_2021}. We cannot assess how the associations we report would differ in a field with a less exclusionary opportunity structure, and a genealogy drawn from a more recent cohort, in which women trumpet players are considerably more present, would be a logical next step for future research.

A further set of limitations concerns the sonic context measure. Because we embed full recordings without source separation, and because VGGish is a general-purpose audio classifier rather than a music-specific model, the measure combines recording technology, production practice, ensemble composition, and instrumentation with the focal player's performance; it does not isolate the trumpet part and cannot be read as a measure of individual tone, articulation, or phrasing. Because each preview samples a recording at a random point, individual excerpts are noisy and may fall on another soloist; the artist-level means, however, are stable, with a split-half procedure recovering the pairwise similarity matrix at a median correlation of $0.73$ ($0.85$ after Spearman--Brown correction; ~\ref{app:splithalf}). This establishes reproducibility rather than construct validity: the stable signal may partly reflect shared repertoire, given how often the jazz canon is re-recorded, or era-specific recording practice; both are components of sonic context as we define it, but neither is evidence of a shared individual style. Future work could attempt to isolate the trumpet part through source separation, which would allow a player's own sound to be distinguished from its sonic context. Aggregation to the artist level also abstracts from within-career change, so the measure captures a player's overall recorded profile rather than their sound at a given moment, a simplification the source itself does not make in depicting Miles Davis as three separate figures. Finally, our collaboration measures do not distinguish the role a musician occupied within an ensemble; since formal and network leadership in jazz groups have different implications for creative outcomes \citep{van2023network}, different kinds of ties may carry different weight in expert judgment.

\subsection{Conclusion}

Despite these limitations, our study offers several insights and a toolkit to examine how cultural fields are conceptualized by experts and participants.

Studies of collaborative creativity focus predominantly on horizontal collaborative ties; our study highlights that paying systematic attention to vertical influence relations is a fruitful extension. Our results show that the structure of expert-recognized influence cannot be explained by considering social and sonic context similarity in isolation. Instead, recognized influence is associated with both dimensions: who a musician played with and the sonic context of the recordings on which that musician appeared. We also find that the association between sonic similarity and recognition weakens as overlap in collaborators increases, and disappears among the pairs with the most shared collaborators. We read this as suggesting that, among musicians who move in the same circles, sounding alike marks membership in a common scene rather than a place in a line of succession. We hope future research can delve deeper into the underlying mechanisms behind the patterns we observe to unearth the conditions under which experts and field participants ascribe influence. The advent of ever-growing datasets capturing behavioral traces of cultural producers, combined with novel techniques for quantifying cultural content, should make it possible to build on our work and conduct similar analyses for other fields, other instruments, and other lineages.

\section*{Declaration of competing interest}
The authors declare that they have no known competing financial interests or personal relationships that could have appeared to influence the work reported in this paper.

\section*{Funding and Acknowledgements}
This research was supported by the Research Excellence Support Fund of Central European University (CEU), administered by the Research Support Office (RSO) [grant number OTHERNSC2525]. This work was also supported by CLARIAH-AT under the project ``Gender, Networks, and Artistic Success: A Computational Analysis of Jazz History''. We also thank Deezer, as this research would not have been possible without the audio files accessed via their API.

\section*{Data availability}
Replication code and the derived matrices are available at \url{https://onurb.codeberg.page/socialstructureofjazz/}.

\appendix
 
\section{Data Coverage}
\label{app:coverage}
 
\begin{figure}[H]
    \centering
    \includegraphics[width=0.88\linewidth]{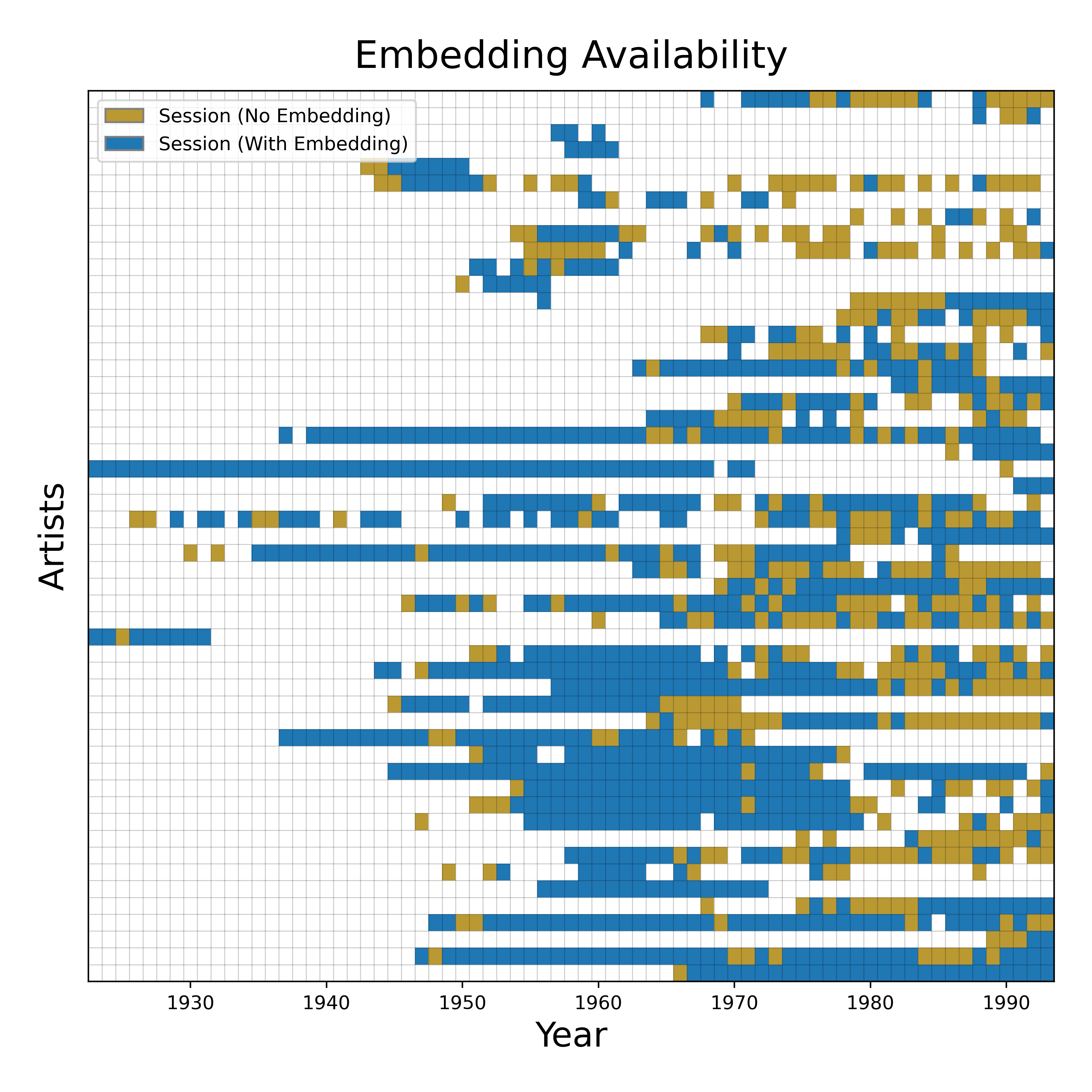}
\caption{\textbf{Temporal coverage of session embeddings.} Availability of
session data across the analyzed period (x-axis) by artist (y-axis). Gold
segments indicate a session record; blue segments indicate sessions for which
audio was retrieved and an embedding generated.}
\label{fig:coverage}
\end{figure}
 
\begin{figure}[H]
    \centering
    \includegraphics[width=0.99\linewidth]{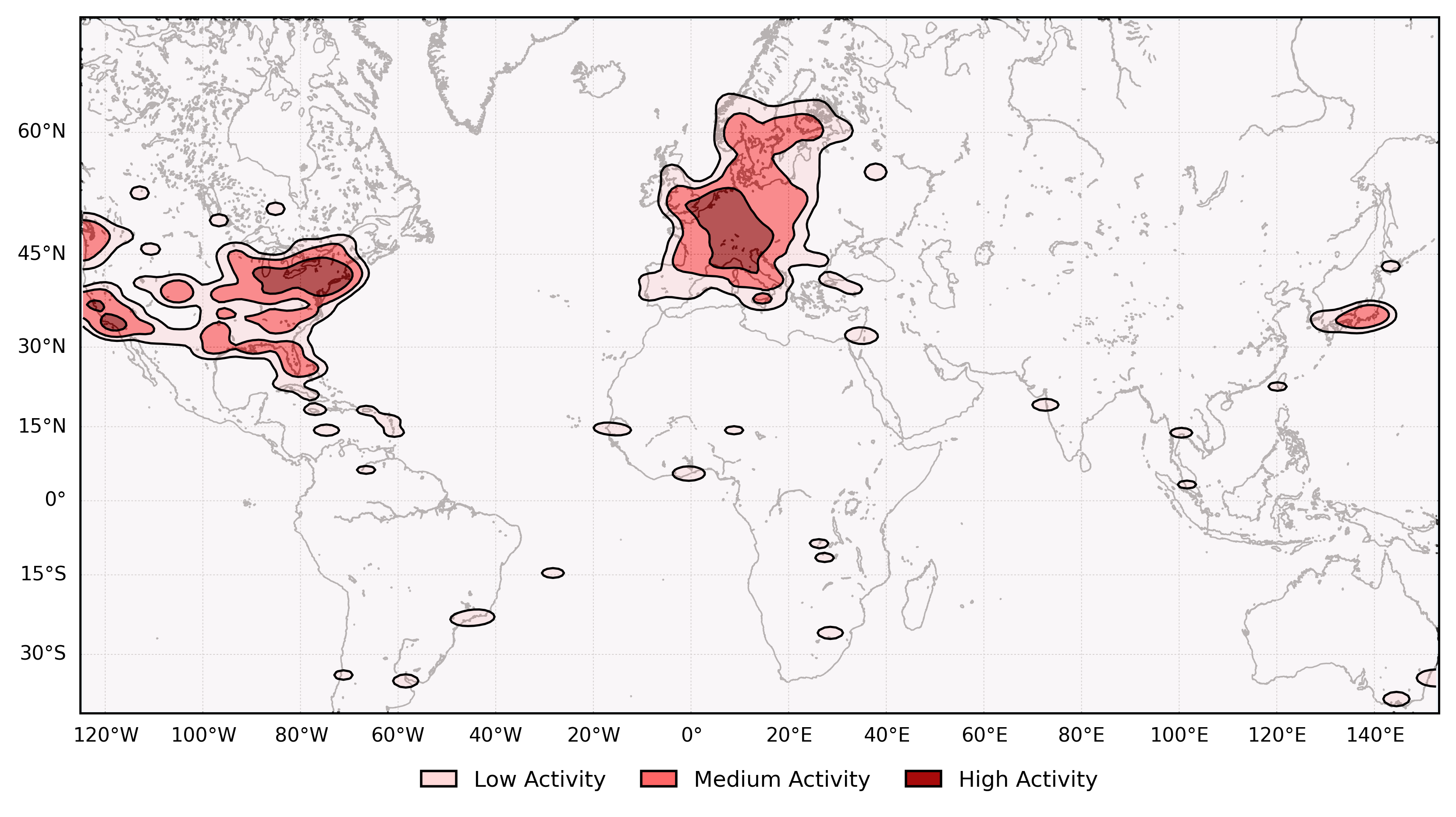}
\caption{\textbf{Spatial distribution of session locations.} Global density of
recording activity based on coordinates retrieved via the Google Places API.
Three primary clusters are visible in the eastern United States, the western
United States, and central Europe, with Japan a secondary hub.}
\label{fig:activity_map}
\end{figure}
 
\section{Structural Properties and the Modelled Space}
\label{app:structure}
 
The genealogy comprises 52 players and 74 directed ties (density 0.028; mean in-
and out-degree 1.42). Table~\ref{tab:netdescriptives} reports its degree
distributions and dyad census.
 
\begin{table}[htbp]
\centering
\caption{Structural properties of the influence network}
\label{tab:netdescriptives}
\small
\begin{tabular}{lr@{\hspace{1.2cm}}lr}
\toprule
\multicolumn{2}{l}{\textit{Out-degree}} & \multicolumn{2}{l}{\textit{In-degree}} \\
0 & 32 & 0 & 3 \\
1 & 8  & 1 & 31 \\
2 & 2  & 2 & 12 \\
3 & 3  & 3 & 5 \\
4 & 1  & 4 & 1 \\
5 & 2  & \multicolumn{2}{l}{} \\
7 & 1  & \multicolumn{2}{l}{} \\
9 & 1  & \multicolumn{2}{l}{} \\
11 & 1 & \multicolumn{2}{l}{} \\
12 & 1 & \multicolumn{2}{l}{} \\
\midrule
\multicolumn{2}{l}{\textit{Dyad census}} & \multicolumn{2}{l}{\textit{Tie orientation by birth year}} \\
Mutual & 2 & Older to younger & 62 \\
Asymmetric & 70 & Younger to older & 8 \\
Null & 1{,}254 & Same birth year & 4 \\
\midrule
\multicolumn{2}{l}{\textit{Edgewise shared partners}} & \multicolumn{2}{l}{} \\
0 & 62 & \multicolumn{2}{l}{} \\
1 & 12 & \multicolumn{2}{l}{} \\
\bottomrule
\end{tabular}
\end{table}
 
Three features bear on model specification. First, the genealogy is not a
directed tree, although it is drawn as a branching diagram. An in-tree requires
every node except the root to have exactly one incoming tie and no cycles;
here 18 players are ascribed two or more predecessors, and two dyads are
reciprocated. Second, and following from the reciprocated dyads, the network is
not a directed acyclic graph. Constraining the ERGM sample space to acyclic or
non-reciprocal configurations would therefore place the observed network outside
the support and leave the likelihood undefined; such a restriction is not
estimable rather than merely worse-fitting. Third, the same applies to
restrictions based on chronological ordering: eight ties run from a younger to
an older player and four connect players born in the same year, for whom no
ordering is defined. No ordering rule is stated in the construction of the
genealogy. We therefore treat all $52 \times 51 = 2{,}652$ ordered pairs as the
modelled space.
 
The distributions in Table~\ref{tab:netdescriptives} also make plain that the
network is sparse and hierarchical: 84$\%$ of ties have no shared partners, most players influence no one, and outgoing ties concentrate in a small number of figures. Sections~\ref{app:jackknife}, \ref{app:qap}, and
\ref{app:gof} report checks addressing whether the reported estimates follow
from this shape rather than from the covariates.
 
\section{Additional Model Specifications}
\label{app:specs}
 
\subsection{Nested model comparison}
 
Neither predictor set is redundant with the other. Adding the collaboration terms to the Sonic Context Model reduces deviance by 26.7 on 4 degrees of freedom ($p < .001$); adding the sonic context terms to the Collab Model reduces it by 35.5 on 3 degrees of freedom ($p < .001$).

\subsection{Direct collaboration and collaboration frequency}
 
Overlap in collaborators and direct co-performance describe different relational
conditions, and we include both. Direct collaboration is a binary indicator of
whether two genealogy players appear together in at least one recording session:
206 of 1{,}326 unordered dyads (15.5\%) meet this condition, against 34 of the 74 influence ties (45.9\%). 
Both measures are positive and significant in the Collab and Full Models (Table~\ref{tab:ergm_models}), and
their correlation of 0.50 indicates that they are related but not
interchangeable.
 
We also examined collaboration frequency, operationalized as the standardized
log count of shared sessions. Entered in place of the binary indicator it is
only marginally associated with ascribed influence ($p = .06$), and it is not
distinguishable from zero once the binary indicator is included. We therefore
retain the binary specification: what predicts ascription is whether two players
recorded together, not how often.

\subsection{Shared predecessors}
 
A shared mentor could in principle account for the reduced ascription of
influence between players with similar sonic contexts. We tested this by adding a
geometrically weighted outgoing shared partner statistic, which counts the
players who influenced both members of a dyad. The term is not distinguishable
from zero ($-0.11$, SE $0.40$) and worsens fit (AIC 390.1 against 389.6 for the reported model), 
while every other estimate is unchanged. Given how few shared
predecessors the genealogy contains (Table~\ref{tab:netdescriptives}), we read
this as evidence that common mentorship does not account for the pattern rather
than as excluding the mechanism: a null estimate on so few configurations has
limited power to rule it out.

\section{Robustness Checks}
\label{app:robustness}
\subsection{Stability of artist-level sonic context profiles}
\label{app:splithalf}

Each 30-second preview samples a recording at a random point, so an individual
excerpt may fall on another soloist or on a passage in which the trumpet is
absent. The relevant question is therefore not whether a single preview is
reliable, which it is not, but whether the artist-level mean computed across
many previews is stable enough to support pairwise comparison.
 
We assess this with a split-half procedure. For each artist we randomly
partition the available previews into two disjoint halves, compute the mean
embedding of each half, and construct the full pairwise similarity matrix from
each set of estimates. The two matrices are then compared on their $1{,}326$
unique off-diagonal entries. Repeating this across 100 random partitions yields
a median correlation of $0.73$ (range $0.62$ to $0.85$), or $0.85$ after the
Spearman--Brown correction for the halved number of previews entering each
estimate. Artists are represented by a median of 47 previews (interquartile
range 20 to 112, minimum 4), so the estimate is weakest for those with the
sparsest coverage.
 
Two points follow. First, the stability is a property of the aggregate rather
than of individual clips: single previews vary considerably, but the mean over a
random half lands close enough to the mean over the other half that the pairwise
similarities are preserved. Second, because the comparison is made on the
similarity matrix rather than on the artist means themselves, the result speaks
to the reproducibility of the dyadic quantity that enters our models, not merely
to the consistency of individual artist positions. The relative positions of
players with respect to one another survive resampling of the underlying audio.

This check concerns reproducibility and not construct validity. By construction, the embeddings do not
isolate the focal musician's playing from recording technology, production,
instrumentation, or ensemble; this composite is what we refer to as sonic
context (Section~\ref{sec:soniccontext}).

\subsection{Threshold sensitivity}
\label{app:thresholds}
 
\begin{figure}[h!]
    \centering
    \includegraphics[width=0.88\linewidth]{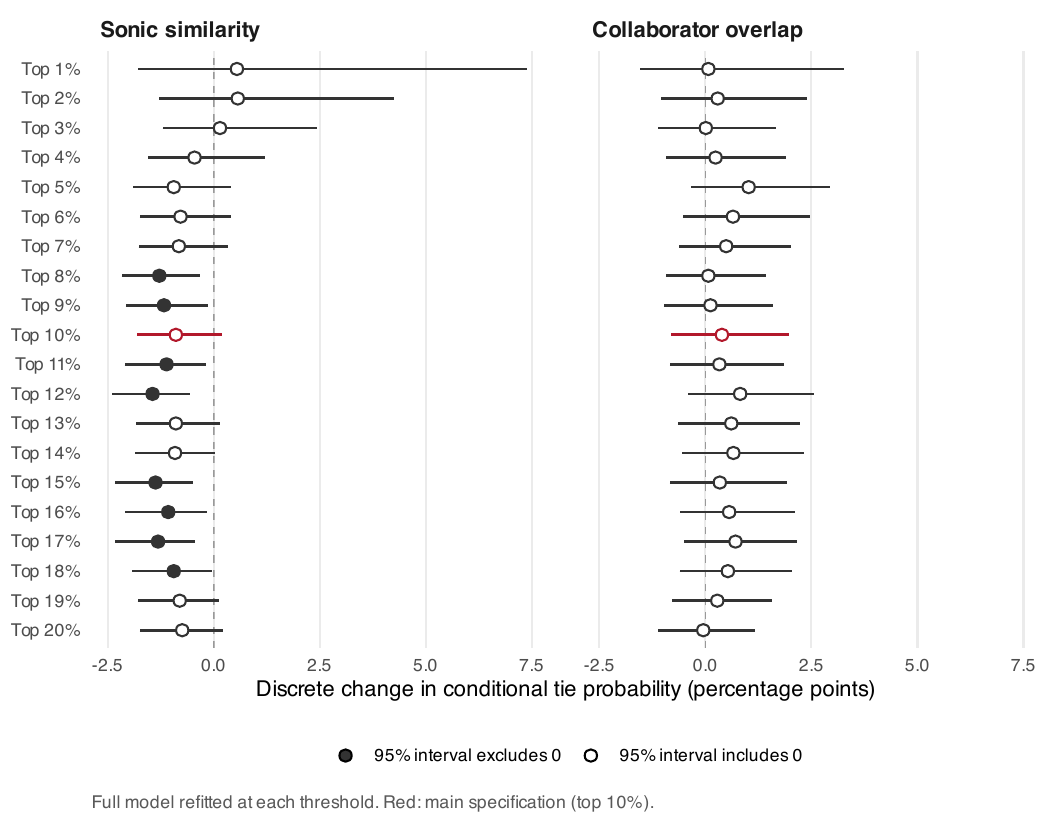}
\caption{\textbf{Sensitivity of the top-share indicators to the threshold definition (full model).} Each row refits the full model with the sonic context similarity (left) and collaborator-overlap (right) indicators coded 1 for dyads in the top share of the respective distribution, from the top 1\% to the top 20\%. Points show the discrete change (0 to 1) in the conditional probability of an ascribed influence tie, in percentage points, with 95\% simulation intervals (2{,}000 draws). Filled points: interval excludes zero. Red: main specification (top 10\%). Because adjacent thresholds share most of their coded dyads, the estimates are not independent.}
\label{fig:AME_audio}
\end{figure}

The top-decile indicators for sonic context similarity and collaboration overlap require
a cutoff. We estimated forty specifications varying this cutoff from the top 1\%
to the top 20\% of each distribution, in baseline and full form.
Figure~\ref{fig:AME_audio} reports average marginal
effects across these specifications. The negative association for sonic context similarity is strongest around the top decile
and does not reach conventional significance at every cutoff; the collaboration indicator is null
throughout. We therefore present the threshold results as exploratory rather
than as establishing a determinate boundary.
 
\subsection{Leave-one-node-out sensitivity}
\label{app:jackknife}
 
Because outgoing ties concentrate in a small number of canonical figures, the
estimates might rest on those players rather than on the structure of the
network as a whole. We assessed this with a node-level jackknife, refitting the
full specification while omitting each of the ten players with the highest
out-degree in turn, together with ten further players drawn at random. Each
replicate omits the focal player and the corresponding row and column of every
dyadic covariate matrix. This follows the logic of jackknife influence
diagnostics \citep{efron_jackknife_1982, cook_detection_1977}, with the node
rather than the observation as the resampling unit
\citep{snijders_non-parametric_1999}. Because every replicate is estimated on a
network of identical size (51 nodes, 2{,}550 ordered dyads) under an identical
specification, coefficients are comparable across fits on the log-odds scale
without rescaling.
 
Every focal estimate retains its sign across all replicates, and the full-sample
estimates fall near the centre of each distribution: sonic context similarity ranges
from 3.35 to 4.45, direct collaboration from 0.95 to 1.30, collaborator overlap
from 0.60 to 1, and the interaction from -1.45 to -0.95. No single player,
including the most central, accounts for any of the reported associations.
 
\subsection{A specification without ERGM structural terms}
\label{app:qap}
 
The reported associations might depend on the ERGM parameterization rather than
on the data, since the geometrically weighted degree terms absorb the
concentration of ties and could in principle carry work attributed to the
dyadic covariates. We therefore re-estimated the full specification as a
multiple regression quadratic assignment procedure with Dekker's double
semi-partialling correction \citep{dekker2007,
krackhardt_predicting_1988}, using 500 permutations. MRQAP does not model endogenous dependence through structural terms; instead, it tests each coefficient against a null distribution obtained by jointly permuting the rows and columns of a matrix, which preserves the dependence structure within that matrix \citep{krackhardt_predicting_1988}. Under double semi-partialling, the permuted matrix is not the dependent network but the residuals of each predictor regressed on all other predictors, which makes the tests robust to collinearity among the predictor matrices \citep{dekker2007}; because these robustness results were established for linear regression, their extension to the logistic case is assumed rather than demonstrated \citep{butts_sna_2008}. Its coefficients are not on a scale comparable to those in
Table~\ref{tab:ergm_models}, so the relevant question is whether signs and
significance survive.
 
They do (Table~\ref{tab:qap}). Sonic context similarity, direct collaboration, and
collaborator overlap each remain positive and significant, the interaction
remains negative, the sonic context threshold term remains negative, and the
collaboration threshold remains null. This is the more demanding comparison for
the interaction and threshold terms in particular, as these are the estimates
most likely to be sensitive to how a model absorbs degree concentration.
 
\begin{table}[htbp]
\centering
\caption{MRQAP estimates (Dekker double semi-partialling, 500 permutations)}
\label{tab:qap}
\small
\begin{tabular}{lrrr}
\toprule
\textbf{Term} & \textbf{Estimate} & \textbf{$\exp(b)$} & \textbf{$p$} \\
\midrule
Intercept                        & $-4.111$ & 0.016  & $<.001$ \\
Overlap time                     & $-0.848$ & 0.428  & $.172$  \\
Overlap space                    & $-0.503$ & 0.605  & $.012$  \\
Sonic context similarity         & $4.139$  & 62.710 & $.002$  \\
Direct collaboration             & $0.878$  & 2.406  & $.010$  \\
Collaborator overlap             & $0.916$  & 2.498  & $.024$  \\
Sonic context $\times$ Collab        & $-1.057$ & 0.348  & $.028$  \\
Sonic context (top decile)       & $-1.179$ & 0.308  & $.026$  \\
Collab (top decile)              & $0.213$  & 1.237  & $.738$  \\
Releases (sender)                & $1.114$  & 3.046  & $.004$  \\
Releases (receiver)              & $0.657$  & 1.930  & $.012$  \\
Releases (absolute difference)   & $-0.357$ & 0.700  & $.222$  \\
Year of birth (sender)           & $-3.260$ & 0.038  & $.122$  \\
Year of birth (receiver)         & $4.128$  & 62.031 & $<.001$ \\
Year of birth (abs.\ difference) & $-4.600$ & 0.010  & $.002$  \\
\midrule
Residual deviance & \multicolumn{3}{r}{438.9 on 2{,}637 df} \\
\bottomrule
\end{tabular}
\begin{tablenotes}
\small
\item Note: two-sided $p$-values from the permutation distribution.
\end{tablenotes}
\end{table}

\subsection{Sonic context similarity and generational proximity}
\label{app:no_time}

Because the audio embeddings partly reflect recording era (Section~\ref{sec:descriptive}), sonic context similarity could act as a proxy for generational proximity rather than as a measure in its own right. To assess this, we re-estimated the full model without the three birth-year terms and temporal overlap, leaving all other terms unchanged.

The coefficient of sonic context similarity changes only slightly, from 4.07 without these controls to 3.81 with them. Sonic context thus carries a small component related to recording era, which the generational terms absorb in the main specification, while the bulk of its association with ascription is distinct from generational proximity. This agrees with the low correlation between sonic context similarity and the absolute difference in birth year. The other focal estimates retain their sign and significance: direct collaboration (0.92 against 1.02), collaborator overlap (1.01 against 0.72), and the interaction ($-1.31$ against $-1.14$).

Omitting the controls worsens fit considerably (AIC 472.7 against 389.6), so the generational terms remain essential to the main specification. Their inclusion ensures that the estimate of sonic context similarity reported in Table~\ref{tab:ergm_models} is net of the era-related component of the embeddings.

\section{Goodness of Fit}
\label{app:gof}

We evaluate fit by comparing the observed network against networks simulated
from the fitted model on statistics not explicitly parameterized in it
\citep{hunter_goodness_2008, lusher_exponential_2013}. Because the genealogy is
directed, we extend the conventional diagnostics to include the full directed
triad census \citep{holland_method_1970} alongside the out- and in-degree
distributions, edgewise shared partners, and geodesic distances. All diagnostics
use 500 simulated networks.

\begin{figure}[!htbp]
    \centering
    \includegraphics[width=0.8\linewidth]{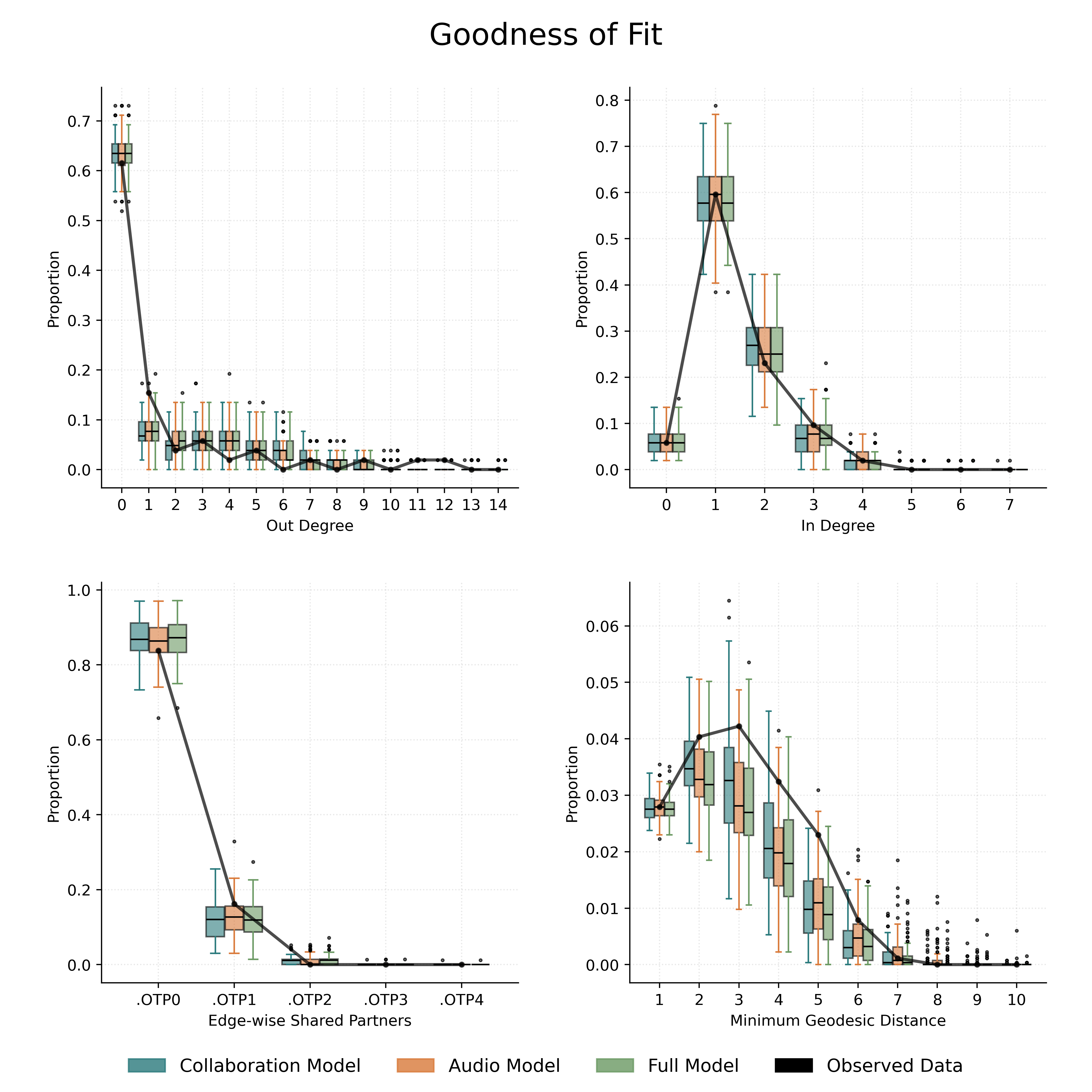}
    \caption{\textbf{Goodness of fit diagnostics.} Structural fit of the Collab,
    Sonic Context, and Full Models across out-degree, in-degree, edgewise shared partners,
    and geodesic distance. Boxes span the interquartile range of 500 simulated
    networks and whiskers extend to the most extreme simulated value within 1.5
    times that range; the solid line marks the observed network.}
    \label{fig:gof_diagnostics}
\end{figure}

The degree and shared partner distributions are reproduced across their support
(Figure~\ref{fig:gof_diagnostics}), with one exception: players holding exactly
one outgoing tie are somewhat more numerous than the model predicts (8 observed
against 3.9 expected, $p = .05$). Fit to the geodesic distribution is less
exact. Observed counts at distances of three, four, and five exceed the
simulated means (112 against 83.3, 86 against 56.9, and 61 against 30.6), and
unreachable pairs fall below it (2{,}188 against 2{,}290.9). No deviation
reaches conventional significance (minimum $p = .08$), but the pattern is
systematic across adjacent distances and indicates that the model
under-produces the long multi-generational chains present in the genealogy. We
report this rather than describing the fit as uniformly close.

The triad census (Table~\ref{tab:triadcensus}) speaks directly to whether the
reported associations follow from the sparse, hierarchical shape of the source
diagram. No triad type deviates significantly from its simulated distribution
(minimum $p = .12$). The model reproduces the near-absence of closure:
transitive triads are rare in both observed and simulated networks (030T: 12
against 7.7), cyclic triads are effectively absent (030C: 0 against 0.3), and
every denser configuration has an observed count of zero against simulated means
below 1. The near triad-free structure of the genealogy is therefore generated by the
estimated degree and covariate effects rather than imposed by the specification.

\FloatBarrier

\begin{table}[!htbp]
\centering
\begin{threeparttable}
\caption{Goodness of fit: directed triad census (500 simulated networks)}
\label{tab:triadcensus}
\small
\begin{tabular}{lrrrr@{\hspace{1.4cm}}lrrrr}
\toprule
\textbf{Type} & \textbf{Obs.} & \textbf{Mean} & \textbf{Max} & \textbf{$p$} &
\textbf{Type} & \textbf{Obs.} & \textbf{Mean} & \textbf{Max} & \textbf{$p$} \\
\midrule
003  & 18{,}832 & 18{,}801.3 & 19{,}388 & .920 & 030T & 12 & 7.7 & 22 & .304 \\
012  & 2{,}872  & 2{,}922.2  & 3{,}627  & .852 & 030C & 0  & 0.3 & 4  & 1.000 \\
102  & 76       & 106.9      & 296      & .508 & 201  & 1  & 0.3 & 5  & .492 \\
021D & 184      & 140.9      & 228      & .124 & 120D & 0  & 0.2 & 4  & 1.000 \\
021U & 16       & 20.6       & 52       & .588 & 120U & 0  & 1.0 & 5  & .872 \\
021C & 85       & 78.4       & 144      & .684 & 120C & 0  & 0.3 & 3  & 1.000 \\
111D & 4        & 2.8        & 13       & .628 & 210  & 0  & 0.1 & 2  & 1.000 \\
111U & 18       & 16.9       & 54       & .920 & 300  & 0  & 0.0 & 0  & 1.000 \\
\bottomrule
\end{tabular}
\begin{tablenotes}
\small
\item \textit{Note:} Triad types follow the Holland--Leinhardt classification.
Monte Carlo $p$-values are two-sided; simulated minima are zero for all types
except 003 (18{,}032), 012 (2{,}379), 021D (74), 021U (7), and 021C (28).
\end{tablenotes}
\end{threeparttable}
\end{table}

\FloatBarrier
\section{Software}
\label{app:software}
 
All models were estimated in \texttt{ergm} \citep{hunter_ergm_2008}; MRQAP
models were estimated in \texttt{sna} \citep{butts_sna_2008}; average marginal
effects were computed with \texttt{ergMargins}
\citep{duxbury_ergmargins_2019}. Random seeds are fixed throughout. Replication
code and the derived matrices are available at \href{https://onurb.codeberg.page/socialstructureofjazz/data.html}{here}.

\clearpage

\bibliography{references, cas-refs}

\end{document}